\PassOptionsToPackage{table}{xcolor}
\documentclass[10pt,a4paper]{article}

\usepackage[a4paper, margin=0.92in]{geometry}
\usepackage[utf8x]{inputenc}
\usepackage{amsmath}
\usepackage{amsthm}
\usepackage{amsfonts}
\usepackage{mathtools}
\usepackage{graphicx}
\usepackage[font={small,it}]{caption}
\usepackage[usenames, dvipsnames]{xcolor}
\usepackage{colortbl}
\usepackage{float}
\usepackage{multirow}
\usepackage{booktabs}
\usepackage{xpatch}
\usepackage{realboxes}
\usepackage{listings}
\definecolor{mygray}{RGB}{232,232,232}
\lstdefinestyle{customc}{
  aboveskip=10pt,
  frame=tlbr,
  framesep=4pt,
  framerule=0pt,
  xleftmargin=9pt,
  xrightmargin=4pt,
  framexleftmargin=5pt,
  language=Matlab,
  basicstyle=\footnotesize\ttfamily,
  morekeywords={integral,arrayfun},
  keywordstyle=\bfseries\color{blue},
  stringstyle=\color{orange},
  backgroundcolor=\color{mygray}
}
\usepackage{array}
\usepackage{tabularx}
\usepackage{extarrows}
\usepackage[affil-it]{authblk}
\usepackage{titlesec}
\usepackage[colorlinks]{hyperref}
\hypersetup{citecolor={blue}}
\usepackage[outdir=./]{epstopdf}
\usepackage[capitalise]{cleveref}
\creflabelformat{equation}{#2#1#3}
\crefname{appsec}{Appendix}{Appendices}
\crefname{equation}{equation}{equations}
\crefname{figure}{Figure}{Figures}
\renewcommand{\P}{\,\mathrm{P}}
\newcommand{\E}{\mathrm{E}}
\newcommand{\Var}{\mathrm{Var}}
\newcommand{\Gumb}{\mathrm{Gumb}}
\newcommand{\Pois}{\mathrm{Pois}}
\newcommand{\tib}{\textcolor{black}}
\newcommand{\ste}{\textcolor{black}}

\titlelabel{\thetitle\quad}

\title{Cancer recurrence times from a branching process model}
\author{Stefano Avanzini and Tibor Antal}
\affil{School of Mathematics, University of Edinburgh, Edinburgh, EH9 3JZ, UK}
\date{}

\begin{document}

\maketitle

\begin{abstract}
{As cancer advances, cells often spread from the primary tumor to other parts of the body and form metastases. This is the main cause of cancer related mortality.}
Here we investigate a conceptually simple model of metastasis formation where metastatic lesions are initiated at a rate which depends on the size of the primary tumor. The evolution of each metastasis is described as an independent branching process. We assume that the primary tumor is resected at a given size and study the earliest time at which any metastasis reaches a minimal detectable size. The parameters of our model are estimated independently for breast, colorectal, headneck, lung and prostate cancers. We use these estimates to compare predictions from our model with values reported in clinical literature. {For some cancer types, we find a remarkably wide range of resection sizes} such that metastases are very likely to be present, but none of them are detectable. Our model predicts that only very early resections can prevent recurrence, and that small delays in the time of surgery can significantly increase the recurrence probability.
\end{abstract}

\tableofcontents



\section{Introduction}
\label{section:introduction}
Metastases develop as cancer cells disseminate from a primary tumor and establish new malignant lesions in the surrounding tissue or at other sites \cite{Sahai:2007}. However, the full process of metastasis formation is much more complex and many related aspects are not yet fully understood. In particular, it is still unclear whether metastases are initiated during early or late stages of carcinogenesis (see e.g.\ \cite{Naxerova:2014,Harper:2016,Reiter:2018}). These details, however, affect the chances of a patient presenting detectable or undetectable metastases at diagnosis, which in turn influence treatment strategies and prognosis. For these reasons, different authors (see e.g.\ \cite{Michor:2006,Haeno:2010} and the references therein) have proposed mathematical models to improve our understanding of the dynamics of metastasis formation.

Metastases frequently arise in cancer patients, and their occurrence greatly diminishes the chances of effective treatment. In fact, even when a therapy is initially successful, metastases often lead to relapse and are responsible for an estimated $90\%$ of cancer related deaths \cite{Chaffer:2011}. Despite this common disease progression, reliable predictions for cancer recurrence rates and times are still lacking \cite{Tsikitis:2014}.

Recently, many generalizations of the Luria-Delbr\"uck model \cite{Luria:1943} have been employed to study specific traits of tumor evolution, such as the development of drug resistance \cite{Iwasa:2006,Komarova:2006,Foo:2013,Bozic:2013}, the role of driver mutations \cite{Durrett:2010a,Durrett:2010b} and metastasis formation \cite{Michor:2006,Haeno:2010,Nicholson:2016,Dingli:2007}. Another line of research focused on temporal features, after the first stochastic model for the time to tumor onset was proposed by Armitage and Doll in their pioneering work on carcinogenesis \cite{Armitage:1954}. A few decades later authors began to investigate stochastic models of tumor latency time. In particular, these works led to mathematical descriptions of optimal schedules of cancer surveillance \cite{Hanin:2001,Hanin:2012}, cure rates \cite{Tsodikov:2003} and cancer recurrence \cite{Yakovlev:1996b}. {These models are studied in the context of survival analysis and reviewed in the excellent book of Yakovlev and Tsodikov \cite{Yakovlev:1996a}.}

In this paper we build a model for cancer recurrence by joining these two approaches, {that is we use Luria-Delbr\"uck type models to study cancer relapse times}. In particular, we consider a deterministically growing tumor seeding metastases at a rate depending on its size \cite{Klein:2009}, and model the evolution of each metastasis (or clone) as independent birth-death branching processes. A similar setup was used by Lea and Coulson to mimic mutations occurring in a growing bacterial population \cite{Lea:1949}. In our model though we interpret these mutation events (from wild-type cells to mutants) as metastasis initiation events. The distribution of mutant close sizes was studied with an exponentially growing wild-type population \cite{Keller:2015} and with more general wild-type growth function \cite{Nicholson:2016}. Kendall \cite{Kendall:1960} also allowed the wild-type population to grow stochastically, but this extension left the mutant behavior unchanged for small initiation (mutation) rates \cite{Kessler:2014,Cheek:2018}. Hence in this paper we model the size of the primary tumor as a deterministic function (focusing on exponential and logistic growth as examples), while allow the seeded metastases to evolve stochastically according to branching processes. 

Within this framework we study the time to cancer relapse, defined as the interval between the primary onset and the first time that any of the metastases reaches a fixed detectable size. Similar characterizations are employed in the threshold models described in \cite{Yakovlev:1996a,Yakovlev:1996b}.

The rest of the paper is organized as follows:
In Results we present our mathematical model of metastases initiation and growth, and derive an explicit formula for the probability distribution of the time to relapse. We then extend our model to include the resection of the primary tumor at a given time and distinguish between synchronous and metachronous metastases.
In Discussion we report parameter estimates for five different cancer types (namely breast, colorectal, headneck, lung and prostate) and analyze the corresponding predictions yielded by our model. Quantitative results are compared with data collected from clinical literature.
In Material and Methods we present details about the mathematical formulation of our model and related derivations.

\section{Results}
\label{section:model}
Our mathematical characterization of the time to cancer recurrence is based on a stochastic model of metastasis formation. We first present the fundamental assumptions and features of this model, and then use them to derive the probability distribution of the time to relapse.

\subsection{Model setup}


\begin{figure}
    \includegraphics[width=16.3cm]{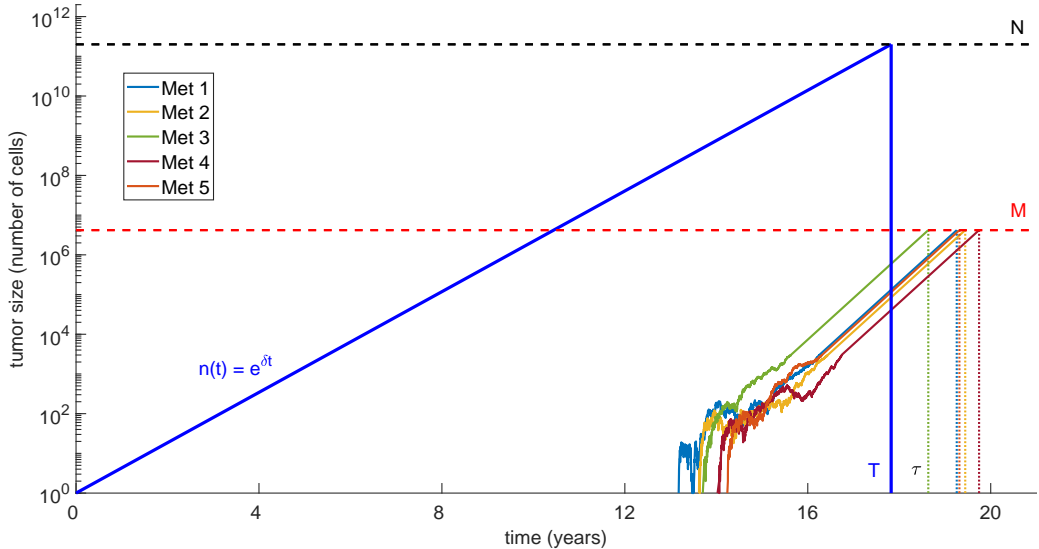}
\caption{{Sample realisation of the model obtained by simulations. The primary tumor grows according to a deterministic exponential function $n(t)$ - depicted by the blue line. It initiates distant metastases at rate $\nu n(t)$, and each of them grows as an independent branching process (only the first five are plotted). The first time $\tau$ that any of these metastases reaches a minimal detectable size $M$ is defined as the time to cancer relapse. Also, the primary tumor is surgically removed at a given time $T$, when it is made of $N=n(T)$ cells. In the realisation shown, the third established metastases (green curve)} {is the first to reach detectable size}, {and hence determines the time to cancer relapse $\tau$. Based on clinical data (summarized in \cref{table:times_size_ranges}), we estimated model parameters (summarized in \cref{table:parameter_estimates}), and here we use those for colorectal cancer, with $N=2\times 10^{11}$.} {Note that a similar illustration for metastasis formation appears in \cite{Tubiana:1986}.}}
\label{figure:illustration}
\end{figure}

We model the number of cells in the primary tumor as a deterministic function of time $n(t)$. The tumor initiates metastases at rate $\nu n(t)$, where $\nu$ is constant. We implicitly assume that all tumor cells can metastasize at the same rate. Since we make no assumptions on $n(t)$, one can define initiation at rate $\nu n(t)^\gamma$ to model scenarios where only a fraction of the primary tumor can metastasize, for example only the cells near its surface or close to blood vessels (see e.g.\ \cite{Haeno:2010}). The initiated metastases are then modelled as independent branching birth{-}death processes \cite{Athreya:2004}, all with the same birth rate $\alpha$ and death rate $\beta$. We assume that they are supercritical, that is they have a positive net growth rate $\lambda=\alpha-\beta>0$, {and consequently grow exponentially for large times \cite{Athreya:2004}. Exponential growth has also been observed in clinical studies \cite{Collins:1956} for untreated human lung metastases, which supports our modelling choice.}

Under these assumptions each metastasis will eventually go extinct with probability $q=\beta/\alpha<1$. The surviving ones instead grow unboundedly and will reach any given size \cite{Athreya:2004}. Let $M$ be a fixed number of cells representing the minimal detectable size of a cancerous lesion. We aim to describe the time to cancer recurrence, defined as the first time $\tau$ that any metastasis reaches the detectable size $M$.

The minimal detectable size $M$ is typically very large, with estimates over $10^6$ (see parameter estimations in Discussion). As the probability that a large supercritical population goes extinct is negligibly small, we assume that each metastasis survives indefinitely if it reaches $M$. Then, due to the splitting property of Poisson processes, the surviving metastases that eventually reach the detectable size are initiated as a non-homogeneous Poisson process $(K_t)_{t\ge0}$ with rate $\nu(1-q)n(t)$. Here $K_t$ denotes the number of metastases established by $t$, conditioned on survival. The expected number of established metastases at time $t$ is thus 
$$
a_t = \E[K_t]=\nu (1-q)\int_0^t n(s) ds
$$
and the probability that at least one is present at $t$ is equal to
\begin{equation}
\label{equation:established_mets}
\P(K_t\ge1)=1-e^{-a_t}
\end{equation}
Surviving metastases are initiated at times
$
\sigma_i:=\inf\{t\ge0:K_t=i\}
$
and are described by i.i.d.\ birth-death processes $(S_i(s))_{s\ge0}$, where $S_i(s)$ is the number of cells in the $i$-th metastasis at time $s$ after its establishment. In particular, we have $S_i(0)=1$ for every $i$. For each of these processes we define
$
\Theta_i:=\inf\{s\ge0:S_i(s)=M\}
$
as the time needed by the $i$-th established metastasis to grow to the detectable size $M$, counting again from its initiation. Since the processes $S_i(s)$ are independent, the hitting times $\Theta_i$ are also independent and identically distributed. As shown in Material and Methods, for $M$ large their distribution asymptotically satisfies
\begin{equation}
\label{equation:gumbel_metastasis}
\P(\Theta_i\le t\mid \Omega_\infty^{(i)})\sim G(t) \equiv e^{-(1-q)Me^{-\lambda t}}
\end{equation}
where $\Omega_\infty^{(i)}$ denotes the eventual survival for the $i$-th metastasis. Interestingly, the distribution $G(t)$ is of a Gumbel type, which generally describes the maximum of independent random variables with exponential {(right)} tail. This Gumbel type has two parameters, $a$ and $b>0$, and distribution function $\exp(-e^{-\frac{x-a}{b}})$. Hence, conditioned on survival, we asymptotically have $\Theta_i \sim \Gumb_\mathrm{max}\left(\frac{\log M(1-q)}{\lambda},\frac{1}{\lambda}\right)$ for every $i$.

\subsection{Time to reach detectable size}
\label{section:time_to_detectability}
Given the definitions in the previous section, we have that the $i$-th metastasis reaches detectable size at time $\tau_i:=\sigma_i+\Theta_i$, measured from primary onset. Metastases are initiated at time $s$ at rate $\nu(1-q)n(s)$ and then reach the detectable size before $t$ with probability $G(t-s)$ \tib{for $s\le t$}. Hence, the thinning property of Poisson processes yields that metastases which become detectable by time $t$ are initiated at time $s$ at rate $\nu(1-q)n(s)G(t-s)$. {Consequently, the number of metastases detectable by $t$ follows a Poisson process $(S_s)_{0\le s\le t}$ with respect to time $s$ for a fixed $t$. In particular,} the number $S_t$ of such metastases established by $t$ is thus a Poisson random variable with mean
\begin{equation}
\label{equation:synchronous_mean}
b_t=\E[S_t]=\nu(1-q)\int_0^t n(s)G(t-s)ds
\end{equation}
The relapse time is defined as the first time that any metastasis reaches the detectable size, $\tau:=\min_i\{\tau_i\}$. Hence, $\tau$ is smaller than $t$ if by that time at least one metastasis that becomes detectable before $t$ is initiated, and so
\begin{equation}
\label{equation:tau_cdf}
\P(\tau\le t)=\P(S_t\ge1)=1-e^{-b_t}
\end{equation}
A sample realisation of our model, including the relapse time $\tau$, is depicted in \cref{figure:illustration}.
In the large detectable size $M$ limit, the relapse time distribution converges to a simpler form (see Material and Methods)
$$
\tau - \frac{1}{\lambda} \log M \xlongrightarrow{d} \overline \tau
$$
where the random variable $\bar{\tau}$ is distributed as
\begin{equation}
\label{equation:tau_scaled_cdf}
\P(\bar{\tau}\le t)=1-\exp\left({-\nu(1-q)\int_{-\infty}^t n(t-s)e^{-(1-q)e^{-\lambda s}}ds}\right)
\end{equation}
Hence for large $M$ the relapse time decomposes as $\tau \approx \frac{1}{\lambda} \log M + \overline \tau$ into a deterministic part which depends only on $\lambda$ and $M$, plus random fluctuations described by $\bar\tau$. This decomposition leads to the estimate $\E[\tau]\sim\frac{1}{\lambda}\log M +C$ for the expected value of the relapse time, where the constant $C=\E[\bar{\tau}]$ can be obtained from \cref{equation:tau_scaled_cdf}.


\begin{figure}
    \includegraphics[width=16.3cm]{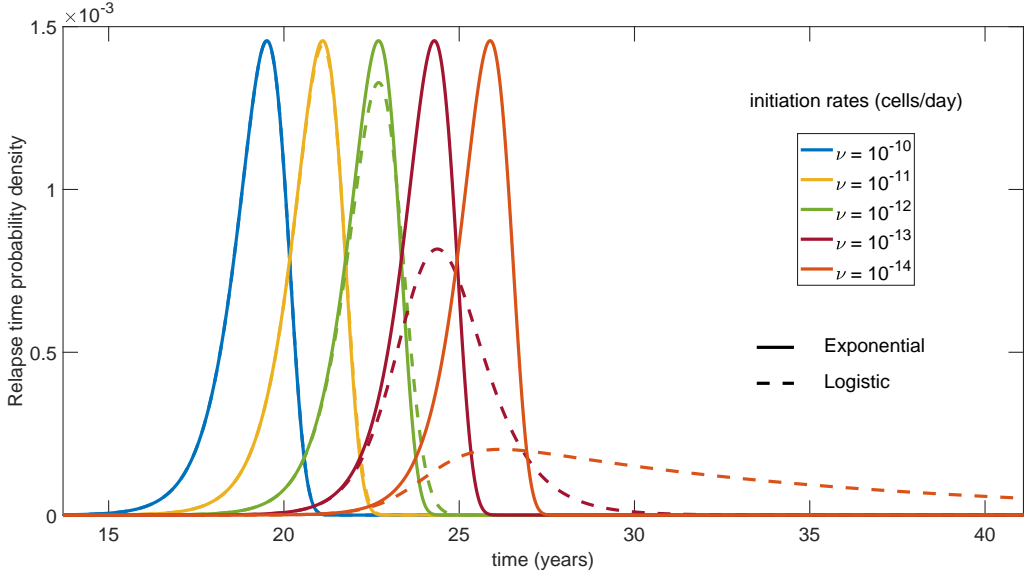}
\caption{Relapse time densities $f_{\tau}(t)=\frac{d}{dt}\P(\tau\le t)$ computed from \cref{equation:tau_cdf} for logistic and exponential primary growths and $\nu=10^{-10},10^{-11},10^{-12},10^{-13},10^{-14}$ cells/day from left to right. Using parameter estimates for colorectal cancer (see \cref{table:parameter_estimates}), the logistic densities (dashed lines) converge to the corresponding exponential ones as the initiation rate increases. Furthermore, in the exponential case and for all the above values of $\nu$, the densities derived from \cref{equation:tau_cdf} and their approximation obtained from \cref{equation:taubar_gumbel_minimum} are indistinguishable.}
\label{figure:tau_densities}
\end{figure}

\subsection{Exponential population growth}
\label{section:exponential_small_nu}
Two commonly employed primary growth functions are the exponential and the logistic ones (see e.g.\ \cite{Preziosi:2003}). These are given by $n(t)=e^{\delta t}$ and $n(t)=\frac{Ke^{\delta t}}{K+e^{\delta t}-1}$, respectively, where $\delta$ denotes the primary tumor net growth rate and $K$ a carrying capacity. Relapse time densities for these two growth types and different initiation rates are shown in \cref{figure:tau_densities}. We observe that as $\nu$ increases, the logistic distributions converge to the exponential ones (see Material and Methods for more details). Moreover, for all our parameter estimates our model predicts the same results with these two growth types. The reason is that the metastases determining the time to relapse are initiated during the early phase of tumor evolution which is almost exponential even for a logistic growth. Therefore, from now on we will focus on exponentially growing primary tumors. {Exponential growth has the additional advantage that if only a portion of primary cells can metastasize and their number is proportional to $n(t)^\gamma$ (say only cells close to the surface of a spherical tumor for $\gamma=2/3$), then this would be equivalent to changing the primary net growth rate, that is using $n(t)=e^{\gamma \delta t}$.} 

\begin{figure}
    \includegraphics[width=16.3cm]{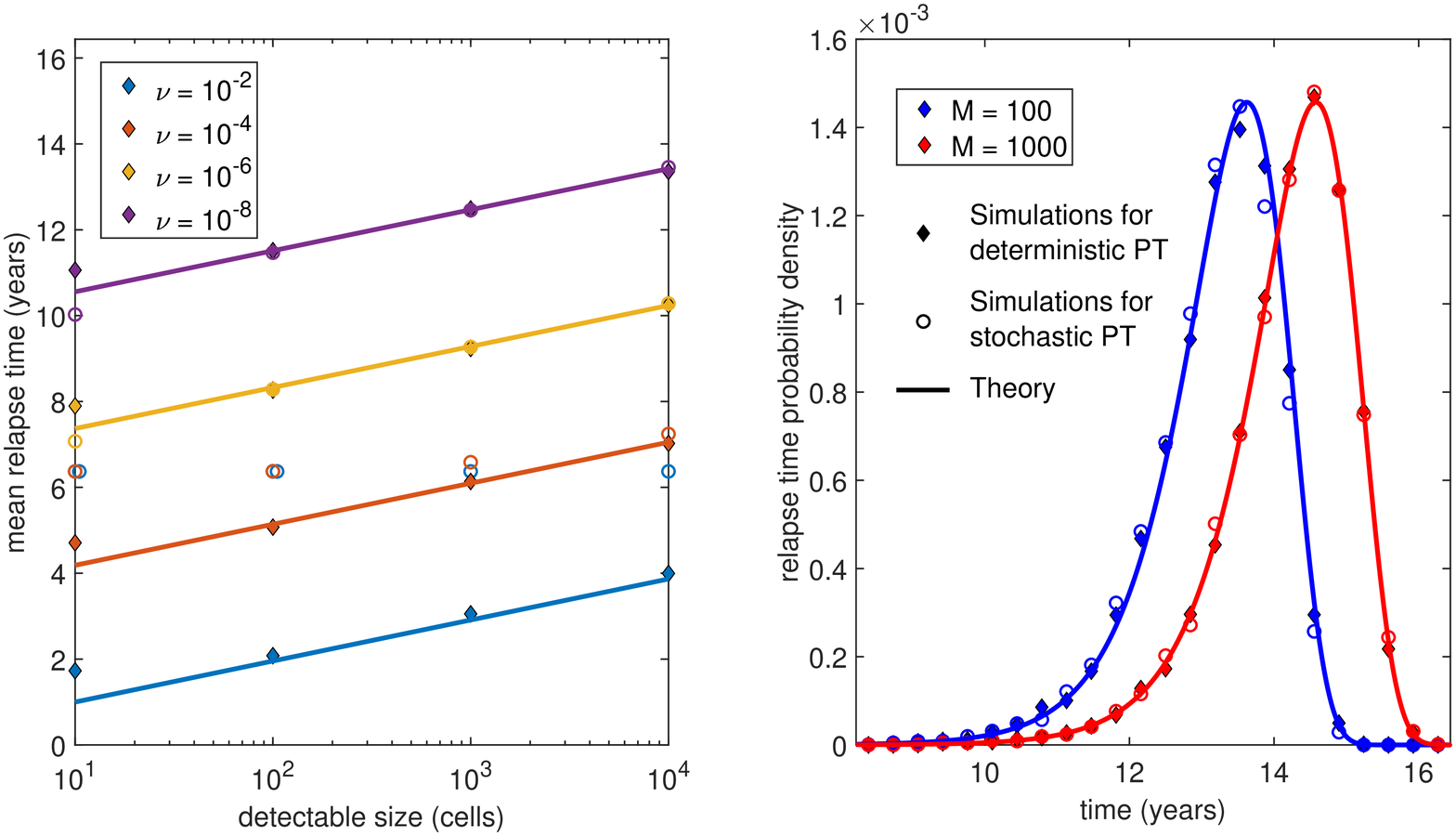}
    \caption{\tib{Relapse time distribution for an exponentially growing primary tumor using parameter estimates for colorectal cancer (see \cref{table:parameter_estimates}). Symbols represent simulation results for a deterministic }\ste{(diamonds)} \tib{or a stochastic primary growth} \ste{(circles, see discussion at the end of the parameter estimation section)}\tib{, while solid lines correspond to the theory in the small $\nu$ - large $M$ limit.} On the left, each starred dot denotes the mean of $1000$ simulations, while lines represent the theoretical expectation given by \cref{equation:tau_expectation_expo_small_nu}. \tib{These match the simulated means well for values of $\nu=10^{-6}$ or less.} On the right, the relapse time densities derived from \cref{equation:gumbel_general} yield a good approximation of the simulated data ($10000$ simulations per curve) for $M=100$ or greater \tib{for both deterministic and stochastic primary growth.}}
\label{figure:tau_expectation_expo}
\end{figure}

Since the initiation rate $\nu$ is {by far the slowest rate in our model}, here we study in detail the most relevant case, that is the small $\nu$ limit for an exponentially growing tumor. The deterministic part of the relapse time remains $\frac{1}{\lambda}\log M$, but interestingly the fluctuations $\bar{\tau}$ are distributed as
\begin{equation}
\label{equation:taubar_gumbel_minimum}
\bar{\tau}\sim\Gumb_\mathrm{min}\left(-\frac{1}{\delta}\log\frac{\nu (1-q)^{1-\frac{\delta}{\lambda}}\Gamma\left(\frac{\delta}{\lambda}\right)}{\lambda},-\frac{1}{\delta}\right)
\end{equation}
This Gumbel distribution describes the minimum of independent random variables with exponential {(left) tail}, has two parameters $a$ and $b<0$ and distribution
$
1-\exp(-e^{-\frac{x-a}{b}})
$.
Parameter $a$ describes a shift in the distribution, and since $a\sim\log\nu$, it explains the equal spacing between the densities in \cref{figure:tau_densities} for logarithmically-spaced values of the initiation rate. Also notice that these curves are left skewed, as it is expected from the Gumbel for the minimum. On the other hand, the Gumbel for the maximum - which describes the fluctuations of the time to detection starting from a single initial cell - is right skewed. 
For small initiation rates $\nu$ and large detectable sizes $M$, the mean relapse time is approximately given by
\begin{equation}
\label{equation:tau_expectation_expo_small_nu}
\E[\tau]\approx \frac{1}{\lambda}\log M+\frac{1}{\delta}\log \frac{\delta}{\nu}+C\,, \qquad C=-\frac{1}{\delta}\left(\log\frac{\delta(1-q)^{1-\frac{\delta}{\lambda}}\Gamma\left(\frac{\delta}{\lambda}\right)}{\lambda}+\gamma_E\right)
\end{equation}
{where $\gamma_E\approx0.5772$ denotes the Euler-Mascheroni constant.} As shown by \Cref{figure:tau_expectation_expo}, this expression fits simulations even for relatively large values of $\nu$ and small values of $M$.


\Cref{equation:tau_expectation_expo_small_nu} highlights a simple dependence of the mean relapse time $E[\tau]$ on $M$ and $\nu$. In Material and Methods we also compute the mean time to detectability of the first established metastasis, $\E[\tau_1]$, where $\tau_1=\sigma_1+\Theta_1$ is equal to the sum of the first initiation time and the hitting time to $M$. Interestingly $\E[\tau_1]$ has the same $M$ and $\nu$ dependence shown in \cref{equation:tau_expectation_expo_small_nu}, but the constant term $\tilde C$ is different. For example, using the parameter estimates for colorectal cancer (see \cref{table:parameter_estimates}) we find $C\approx250$ and $\tilde{C}\approx309$. The reason for this difference is that even in the small $\nu$ - large $M$ limit, later established metastases can outrun the earlier ones in reaching $M$ first.

{While the mean relapse time $\tau$ shows logarithmic increase in terms of $M$ and $\delta/\nu$, its variance stays constant}{, $\Var(\tau)=\pi^2/(6\delta^2)$, see \cref{equation:mean_gumbel}.} {Hence, due to the slow logarithmic growth of the mean, the fluctuations of the relapse time stay relevant even for large detection sizes and small mutation rates.}

\subsection{Relapse time with resection}
Surgery is still the most common and effective type of treatment for solid tumors, although often used in combination with other kind of therapies (see e.g.\ \cite{Bolognese:2009}). However, how the time of resection affects prognosis, and in particular the estimation of the time to relapse, is still unclear. In order to investigate this question in a theoretical framework, we now embed surgery in our model and study how it changes the distribution of the time $\tau$ to relapse. Let us assume that at a given moment after detection a primary solid tumor is surgically removed. This event can be mathematically implemented in our model by considering a resection time $T$ such that $n(t)\equiv0$ for $t\ge T$. In particular, this implies that after $T$ no metastases can be initiated. The number of metastases already established at resection is equal to $K_T$, and their size distribution is given in \cite{Nicholson:2016}. The distribution of the time $\tau$ to relapse can then be expressed exactly as in \cref{equation:tau_cdf}, however here $\tau$ is not a proper random variable. In fact, as $\int_0^\infty n(s)ds = \int_0^{T} n(s)ds<\infty$, there is a positive probability that no metastasis will ever occur (notice that from this point of view our framework can be seen as a cure model - see e.g.\ \cite{Peng:2014}) and in this case we set $\tau=\infty$. The distribution of the relapse time conditioned on at least one metastasis being established by resection is simply
\begin{equation}
\label{equation:tau_cdf_conditional_Kt1}
\P(\tau\le t\mid K_{T}\ge1)=\frac{\P(\tau\le t)}{\P(K_{T}\ge1)}=\frac{1-e^{-b_t}}{1-e^{-a_{T}}}
\end{equation}
{where we used that a relapsing metastasis had to be initiated before resection, that is $\{\tau\le t\}\subset \{K_{T}\ge1\}$.} This conditional distribution for different resection times is depicted in \cref{figure:tau_densities_conditional_Kt1}. In this and following figures, the resection time is shown at the bottom of the figure, and the corresponding resection size $N=e^{\delta T}$ is shown on the top. As $T\to0$ all metastases have to be initiated close to time zero, so the relapse time becomes the time to reach size $M$ from a single cell, which has the Gumbel distribution for the maximum given by \cref{equation:gumbel_metastasis}. If we then increase the resection time, the conditional densities shift to the right by the same amount. Finally, as $T\to\infty$ the relapse time distribution converges to the case without resection
$$
\P(\tau\le t\mid K_{T}\ge 1)\to \P(\tau\le t)
$$
The fluctuations for the unconditional distribution follow a Gumbel type for the minimum, as per \cref{equation:taubar_gumbel_minimum}. Hence, as time increases, the relapse time distribution turns from a right-skewed Gumbel to a left-skewed Gumbel.

Note that the densities in \cref{figure:tau_densities_conditional_Kt1} become indistinguishable from the large time limit as $\P(K_{T}\ge 1)$ approaches one. The reason is that by this time metastases have probably already been initiated and one of the early established ones is likely to relapse first. This suggests that only early enough resection times change the behaviour of the model. For example in the case of colorectal cancer, according to \cref{figure:tau_densities_conditional_Kt1}, only resections of tumors smaller than $10^9$ cells affect the time to recurrence.

Right skewed densities are often chosen to fit probability distributions arising in survival analysis. This is due to the fact that most survival data suffer from right censoring \cite{Allison:2010}, where only a lower bound is known for data points.
Looking at the densities in \cref{figure:tau_densities_conditional_Kt1}, though, we can see both left and right skewed distributions. While a few survival datasets are negatively skewed \cite{Meeker:1998}, cancer relapse times are typically right censored as a consequence of limited follow-up and patients decease before relapse (see e.g.\ \cite{Singh:2011}). However, our model does not take into account any of these events. Furthermore \cite{Hagar:2016} recently proposed a model for the estimation of screening times for colorectal cancer based on the observation that some datasets suffer from left censoring as well.


\begin{figure}
    \includegraphics[width=16.3cm]{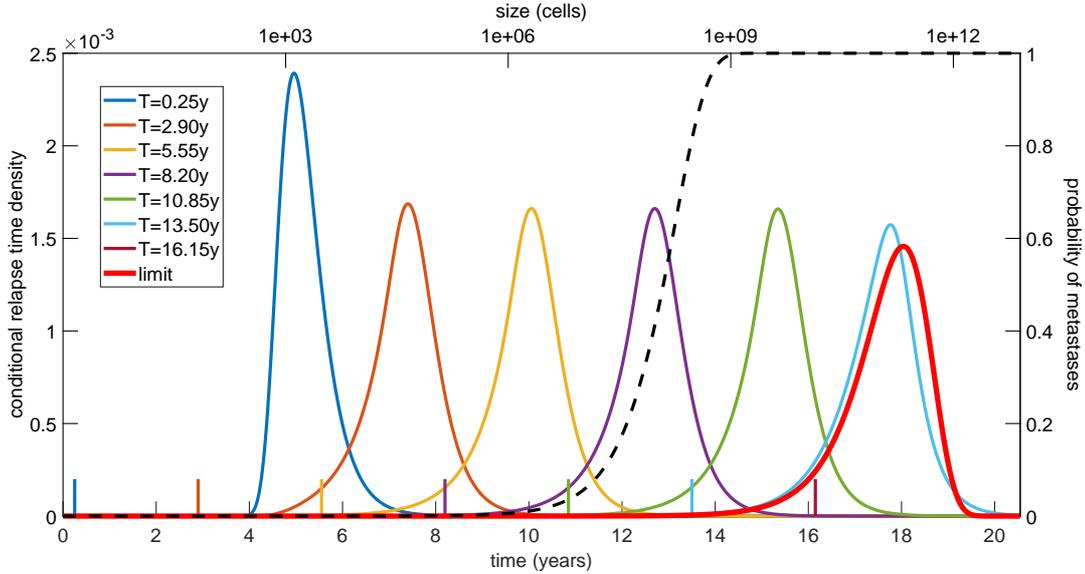}
    \caption{Relapse time densities $f_{\tau}(t\mid K_T\ge1)$ conditioned on at least one metastasis initiated by the time of resection $T$. For different values of $T$, marked with ticks of corresponding colors, these densities are computed by differentiating \cref{equation:tau_cdf_conditional_Kt1}. As $T$ becomes larger, the probability of metastases being established before resection (see \cref{equation:established_mets}) increases and the conditional relapse time densities converge to the red limit one. Here we have used parameter estimates for colorectal cancer (see \cref{table:parameter_estimates}), $n(t)=e^{\delta t}$ and $7$ equally spaced resection times between $0.25$y and $16.15$y. The curves for $T>15$y look identical to the limit density.}
\label{figure:tau_densities_conditional_Kt1}
\end{figure}

\subsection{Metastasis classification}
If the resection is successful and the primary tumor is completely removed, the therapy can still fail due to the formation of metastases. For this reason, it is common practice to start looking for detectable metastases several weeks before the surgery. In this section we thus want to characterize the metastases which are detectable at a given time and those which are not.

In general, {for a fixed time $t$, \tib{the metastasizing process $(K_s)_{0\le s\le t}$ can be split into two independent Poisson processes $(S_s)_{0\le s\le t}$ and $(M_s)_{0\le s\le t}$} describing the initiation of metastases which reach size $M$ before or after $t$}, respectively. Following the same argument we used to derive the relapse time distribution, we obtain that
$$
S_t\sim\Pois\left(b_t\right), \qquad M_t\sim\Pois\left(c_t\right)
$$
where
$$
c_t=a_t-b_t
$$
In particular, we have that the events $\{\tau>t\}$ and $\{S_t=0\}$ are equivalent. We also stress that the definitions above naturally extend to the case of a primary resection, by simply redefining $n(t)$ to be zero after the resection time $T$.

Now, despite an ongoing discussion on the following nomenclature (see e.g.\ \cite{Adam:2015}), in the rest of the paper we will call a metastasis synchronous if it reaches the detectable size $M$ before or up to the time of resection, and metachronous otherwise (hereby the choice of notation $S_t$ and $M_t$). These characterizations immediately allow {us} to estimate the probability of some clinically relevant events. The probability of no synchronous metastases is equal to
\begin{equation}
\label{equation:no_synchronous_mets}
\P(S_T=0)=\P(\tau>T)=e^{-b_T}
\end{equation}
Also, under this condition, relapse is not certain: the probability that at least one metastasis was initiated given that there are no visible ones at resection is
\begin{equation}
\label{equation:metachronous_mets}
P(K_T\ge1\mid S_T=0)=\P(M_T\ge1)=1-e^{-c_T}
\end{equation}
since $S_T$ and $M_T$ are independent. In next section we will study the above and related quantities in greater detail.

\section{Discussion}
\label{section:data}
In this section we compare the predictions provided by our model with clinical data collected for different cancer types. To this purpose, we first need to estimate the parameter values for each of these cancer types.

\subsection{Parameter estimation}
\label{section:parameter_estimation}
The net growth rates of the primary and metastatic tumors, $\delta$ and $\lambda$, are inferred from the corresponding tumor volume doubling times (denoted $DT_{pt}$ and $DT_m$, respectively) as
$$
\delta=\log2/DT_{pt}\,,\quad \lambda=\log2/DT_{m}
$$
These times have been studied by many authors, starting from the influential papers of \cite{Collins:1956,Schwartz:1961,Spratt:1964}. Many authors still refer to these early works, although in some case more recent estimates are available. Colorectal, breast and lung cancers are the most frequently studied. Furthermore, more papers focus on primary doubling times than on metastatic ones.

Similarly, the birth rate $\alpha$ is derived from the potential doubling time $T_{pot}$, defined as the {average} time between cell divisions in the absence of cell death \cite{Jones:2008,Haustermans:1998,Denekamp:1992}. In this case we simply use the estimation 
$$
\alpha = 1/T_{pot}
$$
{Note that some authors (see e.g.\ \cite{Bertuzzi:1997}) define instead $T_{pot}$ as the tumor doubling time in absence of cell death. While in this paper we employ the former definition, the latter would simply yield a factor $\log 2$ in the formula above.}

As for the primary tumor size $N$ at resection, many studies report data on the primary maximum diameter, allowing for ellipsoidal forms. However, given the relatively small tumor volume and the wide interpatient variability, we assume a spherical shape and estimate $d_{pt}$ from the corresponding typical range. By also assuming $10^9$ cells per cm$^3$, the primary size at resection (expressed in number of cells) is thus estimated as $N=\frac{1}{6}\pi d_{pt}^3 10^9$.

\Cref{table:times_size_ranges} summarizes typical ranges of these quantities for five different cancer types, together with the estimates we picked for our model and the corresponding literature references. Difficulties in distinguishing between primary and secondary tumors or in tracking down the primary origin of a metastatic cancer could in principle affect some of these data, but the wide range and multiple references reported reduce the potential impact of this effect.

\begin{table}[H]
\centering
\caption{{Typical ranges of volume doubling times for the primary tumor ($DT_{pt}$) and metastasis ($DT_m$),} {tumor potential doubling time} {($T_{pot}$) and tumor diameter at resection ($d_{pt}$)} for breast, colorectal, headneck, lung and prostate cancer.}
\vspace{5pt}
\setlength{\extrarowheight}{2pt}
\setlength{\belowrulesep}{0pt}
\begin{tabularx}{400pt}{p{64pt}p{57pt}p{70pt}p{45pt}p{110pt}}
 \toprule
 \bf{Cancer type} 	& \bf{Parameter}	& \bf{Typical range} 		& \bf{Estimate}		& \bf{References} \\ 
 \specialrule{0.5pt}{1pt}{0pt} 
 \multirow{4}{*}{Breast} 		& $DT_{pt}$ (days) 		& $103-353$ 			& $210$		& \cite{Fournier:1980,Kuroishi:1990,Peer:1993,Ryu:2014,Foernvik:2015,Zhang:2017} \\ 
						& $DT_{m}$ (days) 		& $85-199$  			& $105$ 		& \cite{Kusama:1972,Friberg:1997} \\ 
						& $T_{pot}$ (days)		& $8-35$  				& $15$		& \cite{Haustermans:1998,Awwad:2013} \\ 
						& $d_{pt}$ (cm)			& $1.4-3$  			& $2.5$cm		& \cite{Zabicki:2006,Lee:2016,DeLAulnoit:2018}\\ 
 \specialrule{0.5pt}{1pt}{0pt} 
 \multirow{4}{*}{Colorectal} 	& $DT_{pt}$ (days) 		& $130-438$ 			& $175$		& \cite{Bolin:1983,Tada:1984,Choi:2013} \\ 
						& $DT_{m}$ (days) 		& $45-155$  			& $105$ 		& \cite{Finlay:1988,Friberg:1997,Tanaka:2004,Tomimaru:2008} \\ 
						& $T_{pot}$ (days)		& $3-4$  				& $4$		& \cite{Wilson:1993,Awwad:2013} \\ 
						& $d_{pt}$ (cm)			& $3.5-5.1$  			& $4.5$cm		& \cite{Bolin:1983,Kornprat:2011,Choi:2013,Ding:2017}\\ 
 \specialrule{0.5pt}{1pt}{0pt} 
 \multirow{4}{*}{Headneck} 	& $DT_{pt}$ (days) 		& $15-256$ 			& $84$		& \cite{Waaijer:2003,Jensen:2007} \\ 
						& $DT_{m}$ (days) 		& $9.5-320$  			& $56$ 		& \cite{Galante:1982,Umino:1997} \\ 
						& $T_{pot}$ (days)		& $1-14$  				& $4$		& \cite{Wilson:1993,Zackrisson:1997} \\ 
						& $d_{pt}$ (cm)			& $1.3-4$				& $2.8$cm		& \cite{Muto:2004,Markou:2011}\\ 
 \specialrule{0.5pt}{1pt}{0pt} 
 \multirow{4}{*}{Lung} 		& $DT_{pt}$ (days) 		& $22-269$ 			& $168$		& \cite{Kerr:1984,Arai:1994,Friberg:1997,Detterbeck:2008,Henschke:2012} \\ 
						& $DT_{m}$ (days) 		& $32-98$  			& $56$ 		& \cite{Spratt:1964,Yoo:2008} \\ 
						& $T_{pot}$ (days)		& $2-17.5$  			& $2.5$		& \cite{Kerr:1984,Fowler:2001} \\ 
						& $d_{pt}$ (cm)			& $1.7-4.1$  			& $2$cm		& \cite{Bando:2002,Strand:2006,Detterbeck:2008}\\ 
 \specialrule{0.5pt}{1pt}{0pt} 
 \multirow{4}{*}{Prostate} 	& $DT_{pt}$ (days) 		& $36-1080$ 			& $392$		& \cite{DAmico:1993,Werahera:2011,Zharinov:2017} \\ 
						& $DT_{m}$ (days) 		& $29-213$  			& $98$ 		& \cite{Berges:1995,Zharinov:2017} \\ 
						& $T_{pot}$ (days)		& $15.2-97.8$	  		& $34$		& \cite{Haustermans:1997,Werahera:2011} \\ 
						& $d_{pt}$ (cm)			& $0.1-2.9$	 	 	& $1.2$cm		& \cite{Renshaw:1999,Johnson:2013}\\ 

\specialrule{1pt}{1pt}{0pt} 
\end{tabularx}
\label{table:times_size_ranges}
\end{table}

Notice that by estimating the rates $\lambda$ and $\alpha$ we also infer values for the death rate $\beta=\alpha-\lambda$ and the extinction probability $q=1-\lambda/\alpha$. For the two remaining parameters, namely the initiation rate $\nu$ and the minimal detectable size of a metastasis $M$, we use common estimates across different cancer types. Various studies report a lowest detectable tumor diameter of $0.2$cm for different cancer types (see e.g.\ \cite{Serres:2012,Fujiwara:2015,Wang:2017}), corresponding to $M\approx4.19\times10^6$ cells. Moreover, several papers argue that the first metastases are likely to be established long before the detection of the primary tumor (see for example \cite{Friberg:1997} and the references therein). In particular, the review of the progression model for metastases formation in \cite{Klein:2009} reports that dissemination starts when the primary diameter is between $0.1$ and $0.4$cm. We thus consider the primary tumor size at the expected time of the first metastasis initiation and estimate it to be $e^{\delta \E[\sigma_1]}=10^8$ cells, corresponding to a diameter of about $0.58$cm. Hence, by using the results in Material and Methods, we set
$$
\nu=\frac{\delta e^{-\gamma_E}}{1-q}e^{-\delta \E[\sigma_1]}
$$
\noindent Finally, the carrying capacity for the logistic primary growth studied in \cref{figure:tau_densities} is set to $K=10^{12}$ \cite{Klein:2009,Chignola:2005}. Overall, we thus found estimates for the following input vector
$$
(DT_{pt},DT_m,T_{pot},d_{pt},d_m,e^{\delta\E[\sigma_1]})
$$
and used them as described above to derive values for our model parameters, i.e.\
$$
(\delta,\lambda,\nu,q,N,M)
$$
Such estimates are summarized in \cref{table:parameter_estimates}.

{Before we proceed to study our model predictions, let us further discuss the assumption of a deterministic primary tumor growth function. Firstly, as we just showed, the only data we found to infer the rate of growth of a primary tumor refer to doubling times, whose notion implicitly assumes an exponential growth. For this reason we focus here on growth functions that (at least in their early stages) show an exponential behaviour.} {Other growth functions, for example $n(t)=ct^3$ for spherically growing tumours or $n(t)=c't^2$ for tumors with active cells only around the surface, could be studied when more data becomes available. Secondly, one could model not just the metastases but also the primary tumor growth as a branching process to account for further stochastic effects.}
{However, due to the large tumor size at resection a branching process model would predict an almost perfect exponential growth around resection time. For this reason we set $n(t)=e^{\delta t}$, which then determines the initial time $t=0$. Note that the tumor is not initiated precisely at $t=0$, but that time is distributed according to a Gumbel distribution, analogously to the results in the Single type process section in Materials and methods. Since the initiation time is not accessible experimentally anyway, for simplicity we use this above definition for $t=0$.}
{In order to justify the exponential deterministic approximation for the primary size, we performed simulations where we modelled} {the primary tumor as a branching process as well. We found that for initiation rates of $\nu=10^{-5}$ or less (and all other parameters set for colorectal cancer) the exponential approximation of the primary causes less than a few percent error in the relapse time distribution \ste{(see \cref{figure:tau_expectation_expo})}. The relationship between stochastic and deterministic wild type populations has been studied rigorously in \cite{Cheek:2018}.}

\begin{table}[H]
\centering
\caption{Parameter estimates for the primary net growth rate $\delta$, the metastatic net growth rate $\lambda$, the initiation rate $\nu$, the extinction probability $q$, the primary tumor size at resection $N$ and the minimal detectable size $M$.}
\vspace{5pt}
\setlength{\extrarowheight}{2pt}
\setlength{\belowrulesep}{0pt}
\begin{tabularx}{400pt}{p{56pt}p{55pt}p{55pt}p{55pt}p{55pt}p{55pt}}
 \toprule
 					& Breast 				& Colorectal 			& Headneck 			& Lung 			& Prostate \\
 \specialrule{0.5pt}{1pt}{0pt} 
$\delta$ (cells/day)		& $0.0033$ 			& $0.0040$ 			& $0.0083$ 			& $0.0041$ 		& $0.0018$ \\
$\lambda$ (cells/day) 	& $0.0066$ 			& $0.0066$ 			& $0.0124$ 			& $0.0124$ 		& $0.0071$ \\
$\nu$ (cells/day) 		& $1.87\times10^{-10}$ 	& $8.42\times10^{-10}$ 	& $9.36\times10^{-10}$ 	& $7.49\times10^{-10}$ 	& $4.13\times10^{-11}$ \\
$q$		 			& $0.9010$ 			& $0.9736$ 			& $0.9505$ 			& $0.9691$ 		& $0.7595$ \\
$N$ (cells) 			& $8.18\times10^9$ 		& $4.77\times10^{10}$	& $1.15\times10^{10}$	& $4.19\times10^{9}$	& $9.05\times10^8$ \\
$M$ (cells) 			& $4.19\times10^6$ 		& $4.19\times10^6$		& $4.19\times10^6$		& $4.19\times10^6$	& $4.19\times10^6$ \\
\specialrule{1pt}{1pt}{0pt} 
\end{tabularx}
\label{table:parameter_estimates}
\end{table}

\subsection{Model predictions}
\label{section:model_predictions}
Now that we have estimated the parameters of our model in \cref{table:parameter_estimates}, we are in position to study its predictions and compare them to clinical data.


\begin{figure}
\includegraphics[width=16.3cm]{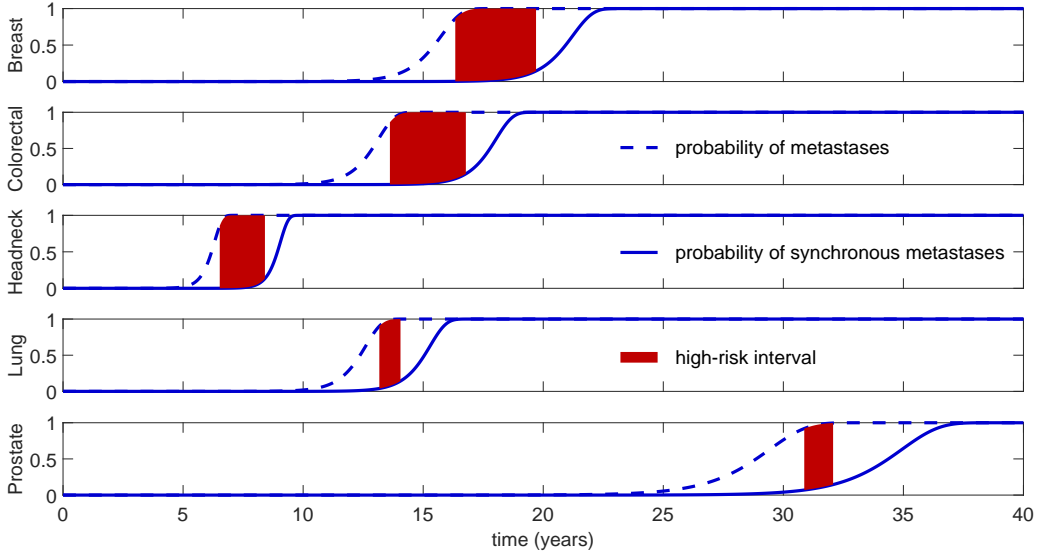}
\caption{Probability of extant metastases, $\P(K_T\ge1)$, dashed curve computed from \cref{equation:established_mets}, and synchronous metastases, $\P(S_T\ge1)$, solid curve computed from \cref{equation:tau_cdf}. These probabilities are plotted as functions of the resection time $T$ for five different cancer types. The primary tumor size at resection is $N=e^{\delta T}$ and thus depends on the primary net growth rate. These resection sizes are discussed in \cref{table:additional_results}. For each cancer type, the shaded areas highlight resection time intervals leading to a probability higher than $85\%$ of established and all undetectable metastases. Using the parameter estimates from \cref{table:parameter_estimates}, the widths of these intervals are $3.41,3.17,1.92,0.94,1.19$ years for breast, colorectal, headneck, lung and prostate cancer respectively.}
\label{figure:high_risk_window_1d}
\end{figure}

Let us start by analyzing the simplest predictions of the model, which are about the presence of synchronous and metachronous metastases. 
\Cref{figure:high_risk_window_1d} shows the probability of initiated metastasis by resection $\P(K_T\ge1)=1-e^{-a_T}$ (\cref{equation:established_mets}), and the probability of visible metastasis by resection $\P(S_T\ge1)=1-e^{-b_T}$ (\cref{equation:tau_cdf}) as functions of the resection time $T$, for five different cancer types. 
{From the figure we notice}
{that, obviously, the probability of having initiated metastasis is always higher than the probability of having visible metastasis at resection.}
For all five cancer types considered, one or more metastases have likely been initiated by the time the primary tumor reaches about $8.2\times10^8$ cells (diameter $1.16$cm). While this value is similar across different primary types (as a consequence of the parameters estimation procedure), the results for the probability of synchronous metastases vary widely. For breast, colorectal, headneck, lung and prostate cancer, \cref{table:additional_results} reports primary tumor sizes at which synchronous metastases might start to appear and are likely to be present, respectively (expressed both in terms of number of cells and tumor diameter). By comparing these values to typical resection sizes in \cref{table:times_size_ranges}, we find that detecting metastases at resection is very likely for lung and prostate cancer and rare for headneck primary tumors.

\begin{table}[H]
\centering
\caption{Resection sizes of the primary tumor which yield a $1\%$ and $99\%$ probability of synchronous metastases, respectively. For each cancer type considered, these sizes are computed with the parameter values in \cref{table:parameter_estimates} and expressed both in terms of number of cells, $N$, and tumor diameter, $d$.}
\vspace{5pt}
\setlength{\extrarowheight}{2pt}
\setlength{\belowrulesep}{0pt}
\begin{tabularx}{400pt}{p{75pt}p{6pt}p{48pt}p{48pt}p{48pt}p{48pt}p{48pt}}
 \toprule
 					& & Breast 				& Colorectal 			& Headneck 			& Lung 			& Prostate \\
 \specialrule{0.5pt}{1pt}{0pt} 
\multirow{2}{*}{$\P(S_T\ge1)>0.01$} 	&$N$ & $1.32\times10^9$		& $2.13\times10^9$		& $7.03\times10^9$		& $1.03\times10^8$	& $6.27\times10^7$ \\
	 				&$d$ & $1.36$ 				& $1.60$	 			& $2.38$	 			& $0.58$	 		& $0.49$ \\
\multirow{2}{*}{$\P(S_T\ge1)>0.99$} 	&$N$ & $6.03\times10^{11}$	& $9.88\times10^{11}$	& $3.22\times10^{12}$	& $4.65\times10^{10}$	& $2.89\times10^{10}$ \\
	 				&$d$ & $10.48$ 				& $12.36$	 			& $18.32$	 			& $4.46$	 		& $3.81$ \\
\specialrule{1pt}{1pt}{0pt} 
\end{tabularx}
\label{table:additional_results}
\end{table}

One of the most challenging scenarios for the development of an effective treatment is when there are only undetectable metastases present. In our framework this scenario corresponds to the event
\begin{equation}
\label{equation:eventundet}
U_T:=\{K_T\ge1,S_T=0\}
\end{equation}
which has probability (see \cref{equation:metachronous_mets,equation:no_synchronous_mets})
\begin{equation}
\begin{split}
\label{equation:extant_all_metachronous}
\P(U_T)&=\P(M_T\ge1,S_T=0)=\P(M_T\ge1)\P(S_T=0)\\
&=e^{-b_T}-e^{-a_T}=\P(K_T\ge1)-\P(S_T\ge1)
\end{split}
\end{equation}
Because of the last identity, the probability of established and all metachronous metastases can be read out from \cref{figure:high_risk_window_1d} as the difference of the two curves. There, the shaded areas highlight intervals of resection times yielding $\P(U_T)>85\%$. These intervals, often referred to as high-risk period \cite{Fillon:2018}, are {especially wide for breast, colorectal and headneck cancers. The reason is that these cancer types have a lower ratio of metastatic over primary net growth rates, so metastases take longer to grow to visible size. Hence, although for these cancer types metastases grow slower, which improves prognosis, they stay undetectable for longer, which poses a challenge for diagnosis.} The estimated resection sizes given in \cref{table:times_size_ranges} fall within or close to these ranges ($\P(U_T)$ equal to $93.87\%,79.83\%,98.35\%,66.04\%$ and $85.85\%$ for the five primary tumor types studied, respectively).
{In general, by assuming that the primary tumor \tib{diameter} at resection fits a normal distribution (with mean computed as the mean of $d_{pt}$ and variance set so that 95\% of the observations belong to its typical range, see Table 1) we estimate that resections for breast and headneck cancers fall in the high-risk window $99.8\%$ and $99.58\%$ of the times respectively, followed by colorectal ($24.67\%$), prostate ($13.41\%$) and lung ($0.69\%$) cancers.}


\begin{figure}
    \includegraphics[width=16.3cm]{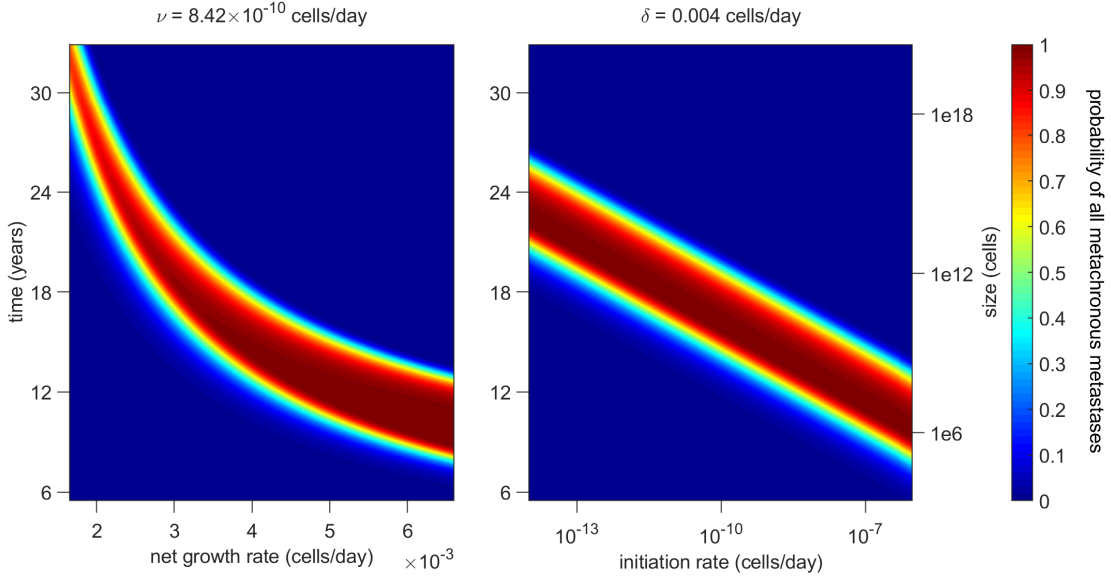}
\caption{Probability of established and all metachronous metastases - $P(U_T)$, as given by \cref{equation:extant_all_metachronous} - plotted as a function of $T$ and $\delta$ (left panel) and of $T$ and $\nu$ (right panel). The parameter estimates used are those for colorectal cancer reported in \cref{table:parameter_estimates}. The plots show that the width of the high-risk interval - the range of resection times such that $\P(U_T)$ is high - stays roughly constant for most parameter values. This width (about $3$ years) shrinks only for metastases growing significantly faster than the primary tumor that initiated them.}
\label{figure:high_risk_window_2d}
\end{figure}

In order to check how robust the presence of a wide high-risk interval is, we plotted in \cref{figure:high_risk_window_2d} the probability of having only undetectable metastasis at detection, $P(U_T)$, for different values of the primary net growth rate $\delta$ and of the initiation rate $\nu$. Other parameters are taken for colorectal cancer. The width of the high-risk interval is constant with respect to $\nu$, and shrinks only as the ratio between the primary and metastatic net growth rate becomes very small. The same qualitative behaviour can be obtained with the parameter estimates for the other cancer types. As most metastases grow up to two times faster than the primary tumor they originated from \cite{Klein:2009}, our model suggests that for a wide choice of parameters there is a substantial range of resection sizes that lead to a high probability of established and all undetectable metastases.

\begin{figure}
    \includegraphics[width=16.3cm]{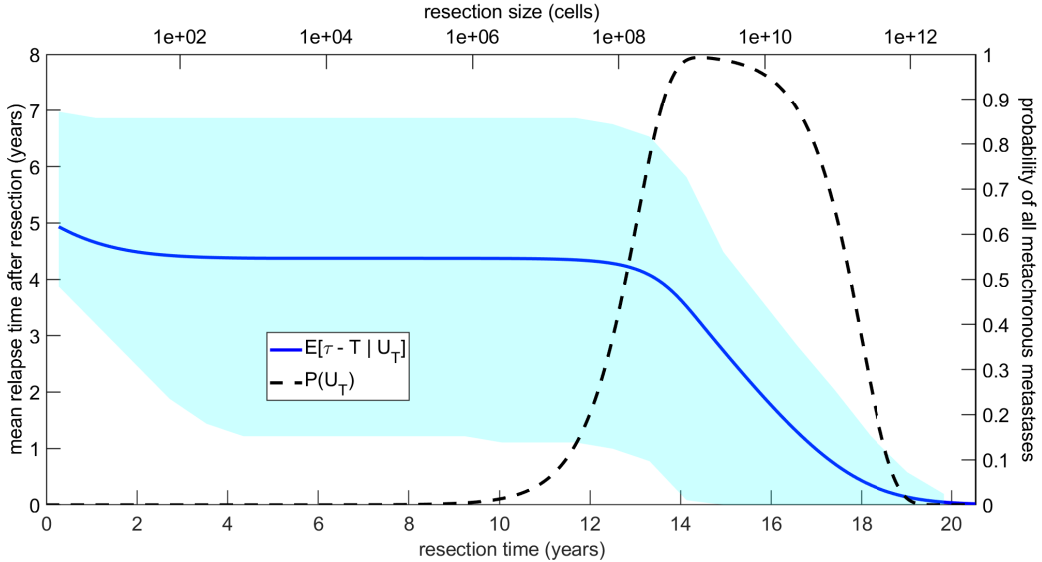}
\caption{Expected relapse time measured from resection, conditioned on extant but all undetectable metastases (blue curve). The dashed line and the light blue shaded area show $\P(U_T)$ and how spread is the conditional relapse time distribution, respectively. The parameter estimates used are those for colorectal cancer reported in \cref{table:parameter_estimates}. For resection times close to zero this conditional expectation coincides with that of the Gumbel distribution given by \cref{equation:gumbel_general}, at about $5$ years. As $T$ starts to increase $\E[\tau-T\mid U_T]$ reflects the convergence highlighted for \cref{figure:tau_densities_conditional_Kt1}, first slightly decreasing and then staying constant around $4.4$ years. Finally, when the resection time falls into the high-risk window, the expected relapse time drops to zero. This suggests that the bigger the primary tumor size is at resection, the faster relapse will occur.}
\label{figure:tau_expectation_after_resection}
\end{figure}

Next, we ask how such a probability, $\P(U_T)$, influences the time to cancer recurrence. The conditional distribution of the relapse time $\tau$ becomes
\begin{equation*}
\P(\tau\le t\mid U_T)=\frac{\P(T<\tau\le t\mid K_T\ge1)}{\P(T<\tau\mid K_T\ge1)}=\frac{e^{-b_T}-e^{-b_t}}{e^{-b_T}-e^{-a_T}}
\end{equation*}
for $t\ge T$, {where we used the definition of $U_T$}
{(see \cref{equation:eventundet})}
{and \cref{equation:tau_cdf_conditional_Kt1}.} From this distribution we compute the expected relapse time measured from resection and conditioned on $U_T$, $\E[\tau-T\mid U_T]$. This expectation and the probability $\P(U_T)$ are plotted in \cref{figure:tau_expectation_after_resection}. We see that for resection sizes smaller than $10^8$ cells the relapse occurs on average between $4$ and $5$ years after resection, independently of the primary size. For resection sizes around $10^8$ cells, undetectable metastases become likely to be present and $\E[\tau-T \mid U_T]$ starts to decrease with tumor size. At about $19$ years the probability of only undetectable metastases present and the conditional mean relapse time both approach zero.

{Let us stress that while some clinical studies report data on the whole distribution of recurrence times, these are usually measured from a varying time of surgery, which corresponds to different primary tumor resection sizes. Therefore, unless the distribution of relapse times is reported together with the corresponding resection sizes, we cannot compare it directly to the predictions of our model. However, we expect the variability of primary sizes at resection to average out across large cohorts of patients, which is why we analyzed the expected value of the time to recurrence.}


\begin{figure}
    \includegraphics[width=16.3cm]{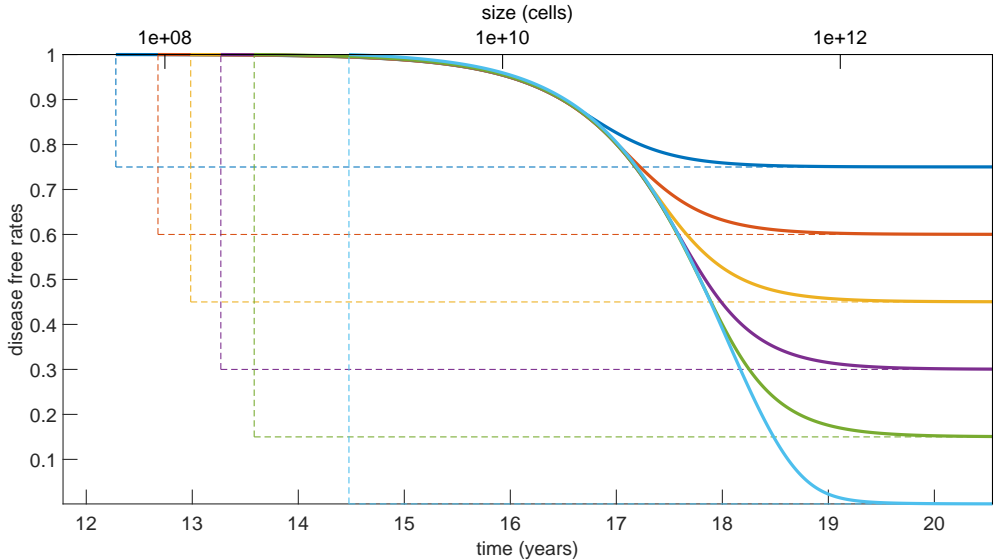}
\caption{Disease-free curves for different resection times. The earlier the primary tumor is resected the higher is the probability that no metastases will arise, or cure probability, represented by the value of the final plateaus. The resection times are chosen so that $\P(K_T=0)=0.75,0.6,0.45,0.3,0.15,0.001$ respectively. With the parameter estimates for colorectal cancer (see \cref{table:parameter_estimates}) these times range from $12.28$ to $14.48$ years, corresponding to sizes between $5.12\times 10^{7}$ and $1.23\times 10^{9}$ cells (diameter $0.46-1.33$cm), respectively.}
\label{figure:disease_free_rates}
\end{figure}

Using the values from \cref{table:parameter_estimates} we tested our model by computing the probability of synchronous metastases and the mean relapse time conditioned on established but all undetectable metastases. The predictions from our model, typical ranges and references for each cancer type considered are summarized in \cref{table:results}. Notice that our predictions for the mean relapse time fall on the lower end of the respective typical ranges. This is expected since we compute the time to recurrence $\tau$ based on the minimal detectable size $M$, while in practice metastases are often detected only at larger sizes. 

In general, for different cancer types it is observed that metastases can grow up to $2$ times faster than the primary tumor they originated from \cite{Klein:2009}, although values as high as $4$ has been proposed \cite{Lee:2006}. Our estimates fall within this range ($\delta/\lambda=4$ for prostate cancer, $3$ for lung and between $1.5$ and $2$ for the others). 

As per the time interval from primary onset to surgery, the typical range is $15-25$ years \cite{Jones:2008}. The high variability in our estimates of $DT_{pt}$ make $T$ fall outside that range for headneck ($T=7.69$y), lung ($T=14.71$y) and prostate ($T=32$y) cancers, classifying the first two as fast growing tumors and the latter as a slow growing one. The singular features that the model predicts for prostate cancer are in accordance with clinical studies (see e.g.\ \cite{Schmid:1993,Berges:1995}).

\begin{table}[H]
\centering
\caption{Typical ranges of $\P(S_T\ge1)$ and $\E[\tau-T\mid U_T]$, predicted value from the model and literature references for each cancer type.}
\vspace{5pt}
\setlength{\extrarowheight}{2pt}
\setlength{\belowrulesep}{0pt}
\begin{tabularx}{400pt}{p{43pt}p{85pt}p{62pt}p{53pt}p{100pt}}
 \toprule
 \bf{Cancer type} 	& \bf{Output}   & \bf{Range from clinical data}  & \bf{Theoretical prediction}   & \bf{Reference} \\ 
 \specialrule{0.5pt}{1pt}{0pt} 
 \multirow{2}{*}{Breast} 		& $P(S_T\ge1)$ ($\%$) 			& $5-10$ 		& $6.13$		& \cite{Andre:2004,Boutros:2015,Yilmaz:2018}\\
						& $\E[\tau-T\mid U_T]$ (days) 		& $590-1022$	& $725$		& \cite{Kim:2013,Fitzpatrick:2013,Nowikiewicz:2015}\\
 \specialrule{0.5pt}{1pt}{0pt} 
 \multirow{2}{*}{Colorectal} 	& $P(S_T\ge1)$ ($\%$) 			& $15-25$		& $20.17$	 	& \cite{Kemeny:2002,Park:2013,Lykoudis:2014,Elferink:2015,Holch:2017,Yilmaz:2018} \\
						& $\E[\tau-T\mid U_T]$ (days) 		& $353-760$	& $356$  		&\cite{Hohenberger:1994,Nordlinger:2009,Elferink:2015,Sturesson:2017,Holch:2017} \\ 
 \specialrule{0.5pt}{1pt}{0pt} 
 \multirow{2}{*}{Headneck} 	& $P(S_T\ge1)$ ($\%$) 			& $1-16.8$	& $1.65$		& \cite{Ferlito:2001,Jain:2013}	 \\
						& $\E[\tau-T\mid U_T]$ (days) 		& $219-623$ 	& $435$		& \cite{Liu:2007,Ebrahimi:2012,Wiegand:2015} \\
 \specialrule{0.5pt}{1pt}{0pt} 
 \multirow{2}{*}{Lung} 		& $P(S_T\ge1)$ ($\%$) 			& $30-55.39$	& $33.96$		& \cite{Yilmaz:2018,Toennies:2014} \\
						& $\E[\tau-T\mid U_T]$ (days) 		& $210-602$	& $249$ 		& \cite{al-Kattan:1997,Hung:2010,Farsi:2017}\\
 \specialrule{0.5pt}{1pt}{0pt} 
 \multirow{2}{*}{Prostate} 	& $P(S_T\ge1)$ ($\%$) 			& $10-34$		& $13.53$		&\cite{Koo:2015,Fontenot:2017,Almeida:2018} \\
						& $\E[\tau-T\mid U_T]$ (days) 		& $730-1131$	& $969$		&\cite{Boorjian:2011,Toussi:2016} \\
\specialrule{1pt}{1pt}{0pt} 
\end{tabularx}
\label{table:results}
\end{table}

The last trait of cancer recurrence that we are going to examine is disease-free rates. These generally correspond to the survival function of the relapse time, $\P(\tau>t)$. However, following the previous discussion we will condition this probability on no synchronous metastases, obtaining
\begin{equation}
\label{equation:tau_cdf_conditional_St0}
\P(\tau\le t\mid S_T=0)
=1-e^{-(b_t-b_T)}
\end{equation}
for $t\ge T$. In this case we do not observe any convergence to the density without resection, because if $T\to 0$ then no metastasis can be initiated and if $T\to\infty$ the condition $S_T=0$ pushes the relapse time to infinity. Let us also stress that our model does not provide information on survival rates, as no modelling of the time to decease is incorporated. Furthermore, notice that $P(\tau>t)$ yields a good description of the disease-free rates in terms of metastases detectability, but not necessarily with respect to cancer symptomaticity.

The {relapse time distribution in case of no synchronous metastasis}, $\P(\tau>t\mid\tau>T)$, for different resection times is shown in \cref{figure:disease_free_rates}, studying again the case of colorectal cancer. As we are not conditioning on at least one metastasis being initiated, there is always a positive probability that relapse will not occur, that is $\tau=\infty$. The resection times are thus chosen so to yield cure probabilities - $\P(K_T=0)$, corresponding to the final plateaus - equal to $0.75,0.6,0.45,0.3,0.15$ and $0.001$, respectively. These times span across a total range of about $2.2$ years. Furthermore, excluding the latest resection time considered, the difference between two consecutive of these $T$ values is between $0.28$ and $0.4$ years. Hence, our model suggests that delays of the order of months in the time of primary resection can lead to a significant decrease in the cure probability.


\begin{figure}
    \includegraphics[width=16.3cm]{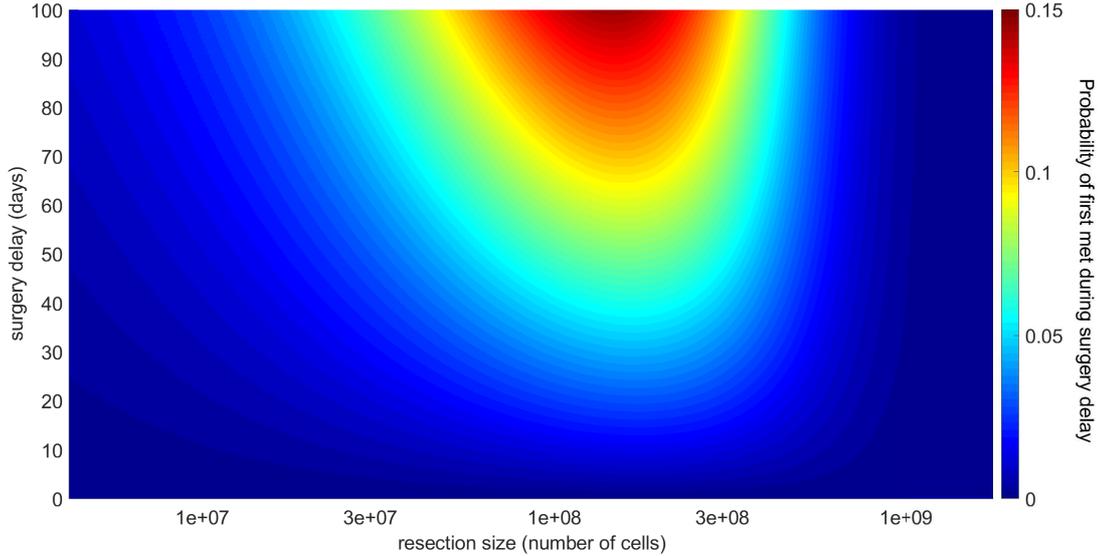}
\caption{{Probability of the first metastasis being initiated during surgery delay. This probability $\P(K_{T+\Delta T}\ge 1, K_{T} = 0)$ - where $T$ is the set time of resection and $\Delta T$ the surgery delay - is plotted as a function of the resection size $N=e^{\delta T}$ (x-axis) and of the delay $\Delta T$ (y-axis). With the parameter estimates for colorectal cancer (see \cref{table:parameter_estimates}) we see that if a primary tumor is resected at a critical size (around $2\times10^8$ cells, diameter $\approx 0.725$cm), surgery delays of 2-3 months can decrease the cure probability of more than $10\%$.}}
\label{figure:surgery_delay}
\end{figure}

{To quantify more precisely the implications of surgery delays, we study the probability that the first metastasis is initiated in the time interval $(T,T+\Delta T)$
$$
 \P(K_{T+\Delta T}\ge 1, K_{T} = 0) = \P(K_{T+\Delta T}\ge 1)-\P(K_{T}\ge 1).
$$ 
This probability is depicted in \cref{figure:surgery_delay} using our parameter estimates for colorectal cancer. We see in the figure that there is a middle range of resection sizes where the recurrence probability can be significantly affected by small surgery delays. For colorectal cancer we estimate that if the primary resection is originally planned for a tumor of diameter between $0.44$ and $0.9$ centimeters ($4.39\times10^7$ and $3.89\times10^8$ cells, respectively), then a surgery delay of 60 days would decrease the cure probability by $5-9\%$. Conversely, tumors smaller than this critical range are less likely to metastasize during the surgery delay,}
{while}
{larger tumors likely metastasized already,}
{so that}
{the effect of surgery delay for these sizes is smaller.}

\subsection{Conclusions}
\label{section:conclusions}
We introduced a model of metastasis formation where metastases are initiated at a time dependent rate, in the simplest case proportional to the size of a growing primary tumor. All initiated metastases then evolve as independent supercritical branching processes. Parameters of the model were estimated for five different cancer types from the clinical literature. We studied the relapse time $\tau$, that is the earliest time when any of the metastases becomes detectable. We obtained the distribution of $\tau$ for a general primary tumor growth and focused in particular on logistic and exponential growth functions. For clinically relevant initiation rates the metastases which relapse first are typically initiated in the early phase of the primary tumor development, which is exponential for both growth functions considered. Hence the distributions of $\tau$ for exponential and logistic primary growths are practically identical unless the initiation rate is unrealistically small ($\nu\approx 10^{-13}$ or smaller) and we can thus exploit the much simpler formulas for the exponentially growing tumor. 

We model the resection of the primary tumor by introducing a cut-off for the growth function $n(t)$. If metastases are likely already established at surgery, their time of relapse is not influenced by the resection timing. We categorized all metastases into synchronous and metachronous and computed corresponding occurrence probabilities. With our estimated parameters we found that the probability of synchronous metastases and the mean relapse time after resection falls in the typical clinical range for all five different cancer types we study.

A challenging scenario for treatment is that of patients with established but all undetectable metastases. For all five cancer types we considered, the probability of this event is high within a significant range of resection sizes. Unfortunately, the typical size of a resected tumor falls in or near this range for all cancer types. We found that relatively small delays in these resection times can cause significant decrease in the cure probability. Within our model, surgery only prevents recurrence if it is done before the onset of the first surviving metastases.

{The parameter estimates summarized in \cref{table:parameter_estimates} yield realistic predictions for several quantities of clinical interest. 
Although in principle we can explore our model predictions across the whole range of parameters, this would often lead to unrealistic outcomes.}
{In this sense the quantitative predictions of our model are quite sensitive to the parameter values, but we have been able to find a combination of parameters that yields realistic results. On the other hand,}
{the qualitative features of our model are more robust to parameter changes, as demonstrated for example in \cref{figure:high_risk_window_2d}.}

{In particular, our estimate for the initiation rate of metastases, $\nu$, is based on the assumption of early dissemination at primary size, $e^{\delta \E[\sigma_1]}=10^8$ cells. However, since the metastatic net growth rate $\delta$ and extinction probability $q$ are estimated independently from data on tumor volume doubling times (see \cref{table:times_size_ranges}), by changing the early dissemination assumption our model predictions could fall outside their typical ranges. For example, for colorectal cancer, assuming}
{$e^{\delta \E[\sigma_1]}=10^9$ cells would lead to the unrealistic values $\E[\tau-T\mid U_T]=836$ days and $P(S_T\ge1)=2.23\%$. Thus, indirectly, our model supports the hypothesis of early metastatic dissemination.}

{Note that in this paper we focused on the presence or absence of synchronous or metachronous metastasis at resection as these events determine if there is ever a relapse ($K_T\ge 1$) or if relapse has already occurred by resection ($S_T\ge 1$). Our model also provides estimates for the number and sizes of metastases at resection, but these are less relevant for the study of the time to cancer relapse, and have already been studied in detail in \cite{Nicholson:2016,Cheek:2018}. A general feature the model predicts is that the cumulative distribution of metastases sizes at resection has a power law tail with exponent $\delta/\lambda$. This power law tail was observed in \cite{Nicholson:2016} using data on 21 patients with colorectal cancer from \cite{Bozic:2013}, and the exponent was found to typically be in the range $0.3-0.8$. Our estimate $\delta/\lambda=0.61$ falls in this range, supporting our parameter inference.}
{The paper above also reports data on the number of visible metastases at surgery. In our model, for a given primary resection size $e^{\delta T}$, this number is a Poisson random variable $S_T$.}
{However, since the primary tumor sizes are not published and likely different for all patients in the data, we could not use this quantity reliably for our parameter estimation. For example, if we infer the initiation rate $\nu$ for colorectal cancer from the probability of visible metastasis at resection (given by $P(S_T\ge 1)=1-e^{-b_T}$ and with estimate $0.2$ from the data reported in \cref{table:results}) we would get essentially the same estimate as in \cref{table:parameter_estimates}. But by using instead the mean number of visible metastases (expressed as $E(S_T|S_T\ge 1)=b_T/(1-e^{-b_T})$, with estimated lower bound $1.4$ from \cite{Holch:2017}), we would infer a 3 times greater estimate for $\nu$. Again, a possible cause of this discrepancy lies in the different resection sizes for patients which we have no data for.}

Metastases are seeded and establish colonies via a specific and complex process called metastatic cascade (for details see e.g.\ \cite{Obenauf:2015}). Since this is known to be a multi-stage process, some authors (see for example \cite{Durrett:2015,Haeno:2010,Haeno:2012} and references therein) have described metastases initiation through two-type stochastic models, where a cell needs to gain the ability to metastasize before it can establish a new metastatic lesion. We did not choose that route for several reasons: (i) the precise details of how and when cells reach this ability are not clear \cite{Jones:2008, Yachida:2010}, (ii) in our model we can think of $n(t)$ as the number of cells which can metastasize and so tailor the two approaches, and (iii) if we assume that an acquired metastatic ability lowers the primary net growth rate {and that the seeding rate is sufficiently small (at most $10^{-4}$ according to simulations),} a branching process model would predict the same exponential growth for the cells with this ability \cite{Cheek:2018, Nicholson:2019}, and hence this would only change the estimate of the initiation rate in our model.

{We did not include into our model a mechanism for metastasis seeding other metastasis, although this phenomena has been observed in clinical studies \cite{Gundem:2015}. The main reason for this omission was the lack of reliable data for the estimation of the secondary seeding rate. By assuming the same primary and secondary seeding rates, however, we would expect metastases to initiate secondary ones when they reach around $10^8$ cells, at which size they are already detectable. Hence, by considering this scenario our predictions for the time to cancer relapse would not change.}

We aim to compare our model in the future to data where relapse times are given jointly with primary tumor sizes at resection. Tumor size is of course not the only relevant factor in predicting relapse times, so the model should be extended to involve other features like a measure of malignancy, possibly as in \cite{Bozic:2010}. Many of the parameters of the model can differ between patients, and also between each metastasis. {Therefore, including a probability distribution for the parameters could also make our model more realistic,}
{provided that such distributions can be estimated from data.}
{Other possible extensions could include interactions among metastatic cells and among metastatic lesions, effects of the immune system, allowing metastases to seed other metastases, and providing an estimate for the fraction of cells which can metastasize, perhaps through modelling angiogenesis.}

\section{Materials and methods}
\label{section:appendix}

In this section we provide more details about the mathematical foundations of our model.

\subsection{Single type process}
\label{section:single_type}
Let $(Z_t)_{t\ge0}$ be a birth-death branching process, i.e.\ a Markov chain on non-negative integers with transition rates
\begin{align*}
i\mapsto
\begin{cases*}
i+1 &\text{ rate } $i\alpha$\\
i-1 &\text{ rate } $i\beta$
\end{cases*}
\end{align*}
The two positive constants $\alpha$ and $\beta$ are called birth and death rate, respectively. In our model we employ this process to describe the evolution of each metastasis. We assume that all metastases have the same birth and death rate and that they are supercritical, that is they have positive net growth rate $\lambda=\alpha-\beta>0$. Moreover, since we only want to model surviving metastasis, we condition on the eventual survival of the process, that is on the event $\Omega_\infty=\{\omega: Z_t(\omega)>0 \text{ for all } t\ge0\}$. The probability of such event is equal to $\P(\Omega_\infty)=1-q$, where $q=\beta/\alpha$ \cite{Athreya:2004}.

We define the first passage time to size $M$ as $T_M:= \inf\{t>0 : Z_t=M\}$. A well known property of branching processes is that $e^{-\lambda t}Z_t \to W$ almost surely as $t\to\infty$, and conditioned on survival and a single initial cell $W\sim \mathrm{Expo}(\lambda/\alpha)$ \cite{Athreya:2004}. Since $W$ and $T_M$ are connected by {$\lim\limits_{M\to\infty} Me^{-\lambda T_M}=W$}, an immediate consequence is that 
\begin{equation}
\label{equation:gumbel_general}
\P(T_M\le t\mid \Omega_\infty \cap \{Z_0=1\})\sim e^{-(1-q)Me^{-\lambda t}}\equiv G(t) \qquad \text{ as }M\to\infty
\end{equation}
Early derivations of this result already appear in \cite{Williams:1965a,Waugh:1972c}. Interestingly, $T_M$ follows the Gumbel distribution $\Gumb_\mathrm{max}\left(\frac{\log M(1-q)}{\lambda},\frac{1}{\lambda}\right)$, where
$$
Y\sim\Gumb_\mathrm{max}(a,b) \quad\Longleftrightarrow\quad \P(Y\le y)=e^{-e^{-\frac{y-a}{b}}}, \quad a\in\mathbb{R},\; b>0
$$
The Gumbel type is an extreme value distribution. If $M_n$ denotes the maximum of $n$ IID random variables $X_i$, the Gumbel distribution above generally describes the limit of $M_n$ as $n\to\infty$, when $X_i$ have an exponential {(right)} tail. A similar definition can be given for the reverse Gumbel distribution, i.e.\ the limit of minimum of IID random variables with an exponential {(left) tail}
$$
Y\sim\Gumb_\mathrm{min}(a,b) \quad\Longleftrightarrow\quad \P(Y\le y)=1-e^{-e^{-\frac{y-a}{b}}}, \quad a\in\mathbb{R},\; b<0
$$
For both of these distributions we have
\begin{equation}
\label{equation:mean_gumbel}
Y\sim \Gumb(a,b) \quad\Longrightarrow\quad \E[Y]=a+b\gamma_E, \quad \Var(Y) = \frac{\pi^2}{6} b^2
\end{equation}
where $\gamma_E\approx0.5772$ denotes the Euler-Mascheroni constant.
{Hence the mean}
{hitting}
{time to $M$ cells grows logarithmically with $M$,}
{while its variance remains constant}
{$$
\E[T_M]=\frac{\log M(1-q)+\gamma_E}{\lambda}, \quad \Var\left(T_M\right)=\frac{\pi^2}{6\lambda^2}
$$
Thus, for sizes $M\approx \alpha/\lambda$ the standard deviation is approximately equal to the mean, but since the mean only grows logarithmically with $M$, fluctuations of $T_M$ stay relevant even for much larger values of $M$.}

\subsection{Scaled relapse time distribution}
\label{section:tau_scaled_cdf_appendix}
In Results we derive the general expression for the relapse time distribution, whose full expression is obtained by combining \cref{equation:gumbel_general,equation:synchronous_mean,equation:tau_cdf}. Here we show how to scale the detectable size $M$ out of this expression, so to split the distribution into a deterministic part and a stochastic term. Let us focus on the integral
$$
\int_0^t n(s)G(t-s)ds=\int_0^t n(s)e^{-(1-q)Me^{-\lambda(t-s)}}ds
$$
and apply the change of variables $z:=t-s-\frac{1}{\lambda}\log M$ to find
$$
\int_{-\frac{1}{\lambda}\log M}^{t-\frac{1}{\lambda}\log M} n\left(t-z-\frac{1}{\lambda}\log M\right)e^{-(1-q)e^{-\lambda z}}dz
$$
By plugging this expression back into \cref{equation:tau_cdf}, at time $t+\frac{1}{\lambda}\log M$ we get
$$
\P\left(\tau-\frac{1}{\lambda}\log M\le t\right)=1-\exp\left(-\nu(1-q)\int_{-\frac{1}{\lambda}\log M}^t n(t-z)e^{-(1-q)e^{-\lambda z}}dz\right)
$$
Hence, as $M$ tends to infinity we obtain 
$$
\tau-\frac{1}{\lambda}\log M \xrightarrow[M\to\infty]{d}\bar{\tau}
$$
where
$$
\P(\bar{\tau}\le t)=1-e^{-\nu(1-q)\int_{-\infty}^t n(t-s)e^{-(1-q)e^{-\lambda s}}ds}
$$
From the last two equations we also see that asymptotically as $M\to\infty$
\begin{equation}
\label{equation:tau_mean}
\E[\tau]\sim\frac{1}{\lambda}\log M+C, 
\qquad C=\E[\bar{\tau}]
\end{equation}

\subsection{Explicit results for exponential primary growth}
\label{section:expo_results_appendix}
Two commonly employed growth functions for primary tumors are the exponential $n_e(t)=e^{\delta t}$ and the logistic $n_l(t)=\frac{K e^{\delta t}}{K+e^{\delta t}-1}$ ones (see e.g.\ \cite{Preziosi:2003}). A logistic growth implies that the primary tumor has a carrying capacity $K$. During the first stages of its development $n_l(t)$ follows the same exponential trajectory of $n_e(t)$ and then approaches a constant as it gets closer to size $K$. As the carrying capacity is typically large, this slowdown for $n_l(t)$ happens around $\hat{t}=\log(K)/\delta$. The differences between the results provided by these two growths functions thus depend on the probability of metastases being initiated by time $\hat{t}$, i.e.\ $\P(K_{\hat{t}}\ge1)\approx1-e^{-\frac{\nu(1-q)K}{\delta}}$. Hence, if
\begin{equation}
\label{equation:expo_logi_difference}
\frac{\nu(1-q)K}{\delta}\gg1
\end{equation}
metastases are likely established in the first stages of the primary growth, i.e.\ when $n_l(t)\approx n_e(t)$. Otherwise, metastases are initiated late in the primary evolution, when the two growth functions are substantially different. This feature is visualized in \cref{figure:tau_densities}, where $\tau$ densities for a logistic growth are shown to converge to the exponential ones as $\nu$ increases and the other parameters are fixed.

Using the parameter values from \cref{table:parameter_estimates}, however, we observe that the condition in \cref{equation:expo_logi_difference} is satisfied for all cancer types considered. In other words, our estimates for $\nu, q, K$ and $\delta$ yield no difference between exponential and logistic growth functions. In light of this, we study in greater detail the results obtained with $n_e(t)$.

\subsubsection{Scaled relapse time}
When $n(t)=n_e(t)=e^{\delta t}$, the relapse time distribution has an expression in terms of special functions. To show this, let us consider the distribution of the scaled relapse time $\overline{\tau}$ as given by \cref{equation:tau_scaled_cdf} and focus on the integral
$$
\int_{-\infty}^t n(t-s)e^{-(1-q)e^{-\lambda s}}ds=e^{\delta t}\int_{-\infty}^t e^{-(1-q)e^{-\lambda s}-\delta s}ds
$$
This can be equivalently written as
$$
e^{\delta t}\int_{-\infty}^t \frac{1}{(1-q)^{\delta/\lambda}}\left[(1-q)e^{-\lambda s}\right]^{\delta/\lambda}e^{-(1-q)e^{-\lambda s}}ds
$$
The last expression then suggests the change of variable $x=(1-q)e^{-\lambda s}$, which leads to
$$
\frac{e^{\delta t}}{\lambda(1-q)^{\delta/\lambda}}\int_{(1-q)e^{-\lambda t}}^\infty x^{\frac{\delta}{\lambda} -1}e^{-x}dx = \frac{e^{\delta t}}{\lambda(1-q)^{\delta/\lambda}}\Gamma\left(\frac{\delta}{\lambda},(1-q)e^{-\lambda t}\right)
$$
and where $\Gamma$ denotes the incomplete upper gamma function $\Gamma(a,t)=\int_t^\infty x^{a-1}e^{-x}dx$. The scaled relapse time distribution for $n(t)=e^{\delta t}$ is thus given by
\begin{equation}
\label{equation:tau_scaled_cdf_expo}
\P(\bar{\tau}\le t)=1-\exp\left(-\frac{\nu (1-q)^{1-\frac{\delta}{\lambda}}e^{\delta t}}{\lambda}\Gamma\left(\frac{\delta}{\lambda},(1-q)e^{-\lambda t}\right)\right)
\end{equation}
Since $\Gamma(1,t)=e^{-t}$, for $\lambda=\delta$ this simplifies to \tib{
$$
\P(\bar{\tau}\le t)=1-\exp\left(-\frac{\nu}{\lambda}e^{-(1-q)e^{-\lambda t}+\lambda t}\right)
$$}

\subsubsection{Small initiation limit}
While the initiation rate can vary significantly across different cancer types, $\nu$ is typically orders of magnitude smaller then all other parameters. Hence, we now investigate $\tau$ distribution in the $\nu\to 0$ limit. Let us first consider the result given by \cref{equation:tau_scaled_cdf_expo} for the scaled time to recurrence $\overline{\tau}$ and write it as
\begin{equation}
\label{equation:tau_scaled_cdf_expo_appendix}
\P(\bar{\tau}\le t)=1-\exp\left(-\frac{\nu (1-q)^{1-\frac{\delta}{\lambda}}\Gamma\left(\frac{\delta}{\lambda}\right)e^{\delta t}}{\lambda}\right)\exp\left(\frac{\nu (1-q)^{1-\frac{\delta}{\lambda}}\phi(t)e^{\delta t}}{\lambda}\right)
\end{equation}
where
$$
 \phi(t)=\int_0^{(1-q)e^{-\lambda t}} s^{\frac{\delta}{\lambda}-1}e^{-s}ds< \int_0^{(1-q)e^{-\lambda t}} s^{\frac{\delta}{\lambda}-1}ds=\frac{\lambda}{\delta}(1-q)^{\frac{\delta}{\lambda}}e^{-\delta t}
$$ 
Notice that the second exponential factor in \cref{equation:tau_scaled_cdf_expo_appendix} is bounded below by $1$ and above by $e^{\frac{\nu(1-q)}{\delta}}$ for all $t\ge0$. Therefore, as $\nu\to0$, the distribution of $\bar{\tau}$ asymptotically converges to
$$
 \P(\bar{\tau}\le t)\sim 1-\exp\left(-\frac{\nu (1-q)^{1-\frac{\delta}{\lambda}}\Gamma\left(\frac{\delta}{\lambda}\right)e^{\delta t}}{\lambda}\right)
$$
Equivalently, for small initiation rates the scaled relapse time $\bar{\tau}$ asymptotically follows a Gumbel distribution for the minimum, 
$
\bar{\tau}\sim\Gumb_\mathrm{min}\left(-\frac{1}{\delta}\log\frac{\nu (1-q)^{1-\frac{\delta}{\lambda}}\Gamma\left(\frac{\delta}{\lambda}\right)}{\lambda},-\frac{1}{\delta}\right)
$.

\subsubsection{Mean relapse time}
By combining the last result with \cref{equation:mean_gumbel,equation:tau_mean} we find that
\begin{equation*}
\E[\tau]\approx \frac{1}{\lambda}\log M+\frac{1}{\delta}\log \frac{\delta}{\nu}+C
\end{equation*}
where $C=-\frac{1}{\delta}\left(\log\frac{\delta(1-q)^{1-\frac{\delta}{\lambda}}\Gamma\left(\frac{\delta}{\lambda}\right)}{\lambda}+\gamma_E\right)$.
Intuitively, the time to relapse is likely to be determined by one of the first established metastases. Given the simple dependence of $\E[\tau]$ on $M$ and $\nu$, we now compare it with the mean time to detectability of the first metastasis, $\E[\tau_1]$. Let us first recall that $\tau_1=\sigma_1+\Theta_1$ is equal to the sum of the first initiation time and the hitting time to $M$. As $\nu\to0$, the distribution of the first arrival $\sigma_1$, given in general by $1-e^{-a_t}$, converges to a reverse Gumbel with parameters $\frac{1}{\delta}\log\frac{\delta}{\nu(1-q)}$ and $-\frac{1}{\delta}$. This implies in particular that 
$$
\E[\sigma_1]=\frac{1}{\delta}\left(\log\frac{\delta}{\nu(1-q)}-\gamma_E\right)=\frac{1}{\delta}\log\frac{\delta}{\nu}+C_1
$$
where $C_1=-\frac{\log(1-q)+\gamma_E}{\delta}$.
Moreover, the hitting times $\Theta_i$ follow the Gumbel distribution $G(t)$ - see \cref{equation:gumbel_metastasis} - and hence
$$
\E[\Theta_i]=\frac{1}{\lambda}\log M-C_2
$$
for every $i$, where $C_2=-\frac{\log(1-q)+\gamma_E}{\lambda}$. Joining the last two results we get
\begin{equation}
\label{equation:tau1_expectation_expo_small_nu}
\E[\tau_1]=\E[\sigma_1+\Theta_1]=\frac{1}{\lambda}\log M+\frac{1}{\delta}\log\frac{\delta}{\nu}+\tilde{C}
\end{equation}
where $\tilde{C}=C_1-C_2$. By comparing \cref{equation:tau1_expectation_expo_small_nu} with the expression for $\E[\tau]$, we notice indeed the same $M$ and $\nu$ dependence, but the constants $C$ and $\tilde{C}$ have different analytical forms.

\subsubsection{Numerical computation}
Finally, all the plots and computations reported in this paper have been performed on Matlab R2018b. The lines of code below provide an efficient way (in the example for the exponential case) to calculate the relapse time distribution given by \cref{equation:tau_cdf} for a vector of times \lstinline[style=customc]{tspan}.
\begin{lstlisting}[language=Matlab,style=customc]
n = @(t)(exp(delta*t));
G = @(t)(exp(-(1-q)*M*exp(-lambda*t)));
F = @(t)(1-exp(-nu*(1-q)*integral(@(s)(n(s).*G(t-s)),0,t,'ArrayValued',true)));
x = arrayfun(@(t)F(t),tspan);
\end{lstlisting}

\section*{Acknowledgments}
We thank Ivana Bozic, David Cheek, Jasmine Foo, Kevin Leder, Michael Nicholson and Johannes Reiter for helpful discussions.


%
%
%


\begin{thebibliography}{100}

\bibitem{Sahai:2007}
Sahai E.
\newblock Illuminating the metastatic process.
\newblock Nature Reviews Cancer. 2007;7(10):737--749.
\newblock doi:{10.1038/nrc2229}.

\bibitem{Naxerova:2014}
Naxerova K, Brachtel E, Salk JJ, Seese AM, Power K, Abbasi B, et~al.
\newblock Hypermutable {DNA} chronicles the evolution of human colon cancer.
\newblock Proceedings of the National Academy of Sciences.
  2014;111(18):E1889--E1898.
\newblock doi:{10.1073/pnas.1400179111}.

\bibitem{Harper:2016}
Harper KL, Sosa MS, Entenberg D, Hosseini H, Cheung JF, Nobre R, et~al.
\newblock Mechanism of early dissemination and metastasis in Her2$+$ mammary
  cancer.
\newblock Nature. 2016;540(7634):588--592.
\newblock doi:{10.1038/nature20609}.

\bibitem{Reiter:2018}
Reiter JG, Makohon-Moore AP, Gerold JM, Heyde A, Attiyeh MA, Kohutek ZA, et~al.
\newblock Minimal functional driver gene heterogeneity among untreated
  metastases.
\newblock Science. 2018;361(6406):1033--1037.
\newblock doi:{10.1126/science.aat7171}.

\bibitem{Michor:2006}
Michor F, Nowak MA, Iwasa Y.
\newblock Stochastic dynamics of metastasis formation.
\newblock Journal of Theoretical Biology. 2006;240(4):521--530.
\newblock doi:{10.1016/j.jtbi.2005.10.021}.

\bibitem{Haeno:2010}
Haeno H, Michor F.
\newblock The evolution of tumor metastases during clonal expansion.
\newblock Journal of Theoretical Biology. 2010;263(1):30--44.
\newblock doi:{10.1016/j.jtbi.2009.11.005}.

\bibitem{Chaffer:2011}
Chaffer CL, Weinberg RA.
\newblock A Perspective on Cancer Cell Metastasis.
\newblock Science. 2011;331(6024):1559--1564.
\newblock doi:{10.1126/science.1203543}.

\bibitem{Tsikitis:2014}
Tsikitis VL, Larson DW, Huebner M, Lohse CM, Thompson PA.
\newblock Predictors of recurrence free survival for patients with stage {II}
  and {III} colon cancer.
\newblock {BMC} Cancer. 2014;14(1).
\newblock doi:{10.1186/1471-2407-14-336}.

\bibitem{Luria:1943}
Luria SE, Delbr\"uck M.
\newblock Mutations of bacteria from virus sensitivity to virus resistance.
\newblock Genetics. 1943;48(6):491--511.

\bibitem{Iwasa:2006}
Iwasa Y, Nowak MA, Michor F.
\newblock Evolution of Resistance During Clonal Expansion.
\newblock Genetics. 2006;172(4):2557--2566.

\bibitem{Komarova:2006}
Komarova N.
\newblock Stochastic modeling of drug resistance in cancer.
\newblock Journal of Theoretical Biology. 2006;239:351--366.

\bibitem{Foo:2013}
Foo J, Leder K.
\newblock Dynamics of cancer recurrence.
\newblock The Annals of Applied Probability. 2013;23(4):1437--1468.
\newblock doi:{10.1214/12-AAP876}.

\bibitem{Bozic:2013}
Bozic I, Reiter JG, Allen B, Antal T, Chatterjee K, Shah P, et~al.
\newblock Evolutionary dynamics of cancer in response to targeted combination
  therapy.
\newblock eLife. 2013;2.

\bibitem{Durrett:2010a}
Durrett R, Moseley S.
\newblock Evolution of resistance and progression to disease during clonal
  expansion of cancer.
\newblock Theoretical Population Biology. 2010;77:42--48.

\bibitem{Durrett:2010b}
Durrett R, Foo J, Leder K, Mayberry J, Michor F.
\newblock Evolutionary dynamics of tumor progression with random fitness
  values.
\newblock Theoretical Population Biology. 2010;78(1):54--66.
\newblock doi:{10.1016/j.tpb.2010.05.001}.

\bibitem{Nicholson:2016}
Nicholson MD, Antal T.
\newblock Universal Asymptotic Clone Size Distribution for General Population
  Growth.
\newblock Bulletin of Mathematical Biology. 2016;78(11):2243--2276.
\newblock doi:{10.1007/s11538-016-0221-x}.

\bibitem{Dingli:2007}
Dingli D, Michor F, Antal T, Pacheco JM.
\newblock The emergence of tumor metastases.
\newblock Cancer Biology {\&} Therapy. 2007;6(3):383--390.
\newblock doi:{10.4161/cbt.6.3.3720}.

\bibitem{Armitage:1954}
Armitage P, Doll R. The Age Distribution of Cancer and a Multi-stage Theory of
  Carcinogenesis; 1954.

\bibitem{Hanin:2001}
Hanin LG, Tsodikov AD, Yakovlev AY.
\newblock Optimal schedules of cancer surveillance and tumor size at detection.
\newblock Mathematical and Computer Modelling. 2001;33(12-13):1419--1430.
\newblock doi:{10.1016/s0895-7177(01)80023-6}.

\bibitem{Hanin:2012}
Hanin L, Pavlova L.
\newblock Optimal screening schedules for prevention of metastatic cancer.
\newblock Statistics in Medicine. 2012;32(2):206--219.
\newblock doi:{10.1002/sim.5474}.

\bibitem{Tsodikov:2003}
Tsodikov AD, Ibrahim JG, Yakovlev AY.
\newblock Estimating Cure Rates From Survival Data.
\newblock Journal of the American Statistical Association.
  2003;98(464):1063--1078.
\newblock doi:{10.1198/01622145030000001007}.

\bibitem{Yakovlev:1996b}
Yakovlev AY.
\newblock Threshold models of tumor recurrence.
\newblock Mathematical and Computer Modelling. 1996;23(6):153--164.
\newblock doi:{10.1016/0895-7177(96)00024-6}.

\bibitem{Yakovlev:1996a}
Yakovlev AY, Tsodikov AD, Asselain B.
\newblock Stochastic Models of Tumor Latency and Their Biostatistical
  Applications.
\newblock World Scientific; 1996.

\bibitem{Klein:2009}
Klein CA.
\newblock Parallel progression of primary tumours and metastases.
\newblock Nature Reviews Cancer. 2009;9:302.

\bibitem{Lea:1949}
Lea DE, Coulson CA.
\newblock The distribution of the numbers of mutants in bacterial populations.
\newblock Journal of Genetics. 1949;49(3):264--285.
\newblock doi:{10.1007/BF02986080}.

\bibitem{Keller:2015}
Keller P, Antal T.
\newblock Mutant number distribution in an exponentially growing population.
\newblock Journal of Statistical Mechanics: Theory and Experiment.
  2015;2015(1).

\bibitem{Kendall:1960}
Kendall DG.
\newblock Birth-and-death processes, and the theory of carcinogenesis.
\newblock Biometrika. 1960;47:13--21.

\bibitem{Kessler:2014}
Kessler DA, Levine H.
\newblock Scaling Solution in the Large Population Limit of the General
  Asymmetric Stochastic Luria{\textendash}Delbr\"uck Evolution Process.
\newblock Journal of Statistical Physics. 2014;158(4):783--805.
\newblock doi:{10.1007/s10955-014-1143-3}.

\bibitem{Cheek:2018}
Cheek D, Antal T.
\newblock Mutation frequencies in a birth{\textendash}death branching process.
\newblock The Annals of Applied Probability. 2018;28(6):3922--3947.
\newblock doi:{10.1214/18-aap1413}.

\bibitem{Tubiana:1986}
Tubiana M.
\newblock The growth and progression of human tumors: Implications for
  management strategy.
\newblock Radiotherapy and Oncology. 1986;6(3):167--184.
\newblock doi:{10.1016/s0167-8140(86)80151-7}.

\bibitem{Athreya:2004}
Athreya KB, Ney PE.
\newblock Branching Processes.
\newblock Dover Publications; 2004.


\bibitem{Collins:1956}
Collins VP, Loeffler RK, Tivery H.
\newblock Observations on growth rates of human tumors.
\newblock The American journal of roentgenology, radium therapy, and nuclear
  medicine. 1956;76:988--1000.
  
\bibitem{Preziosi:2003}
Preziosi L.
\newblock Cancer Modelling and Simulation.
\newblock Chapman \& Hall/CRC Mathematical and Computational Biology. CRC
  Press, Taylor \& Francis Group; 2003.

\bibitem{Bolognese:2009}
Bolognese A, Izzo L.
\newblock Surgery in Multimodal Management of Solid Tumors.
\newblock Springer Milan; 2009.

\bibitem{Peng:2014}
Peng Y, Taylor JMG.
\newblock Cure Models.
\newblock In: Handbook of Survival Analysis. Chapman \& Hall; 2014. p.
  113--134.

\bibitem{Allison:2010}
Allison PD.
\newblock Survival analysis using SAS: a practical guide.
\newblock 2nd ed. SAS Publishing; 2010.

\bibitem{Meeker:1998}
Meeker WQ, Escobar LA.
\newblock Statistical Methods for Reliability Data.
\newblock John Wiley \& Sons Inc.; 1998.

\bibitem{Singh:2011}
Singh R, Mukhopadhyay K.
\newblock Survival analysis in clinical trials: Basics and must know areas.
\newblock Perspectives in Clinical Research. 2011;2(4):145.
\newblock doi:{10.4103/2229-3485.86872}.

\bibitem{Hagar:2016}
Hagar YC, Harvey DJ, Beckett LA.
\newblock A multivariate cure model for left-censored and right-censored data
  with application to colorectal cancer screening patterns.
\newblock Statistics in medicine. 2016;35:3347--3367.

\bibitem{Adam:2015}
Adam R, de~Gramont A, Figueras J, Kokudo N, Kunstlinger F, Loyer E, et~al.
\newblock Managing synchronous liver metastases from colorectal cancer: A
  multidisciplinary international consensus.
\newblock Cancer Treatment Reviews. 2015;41(9):729--741.
\newblock doi:{10.1016/j.ctrv.2015.06.006}.

\bibitem{Schwartz:1961}
Schwartz M.
\newblock A biomathematical approach to clinical tumor growth.
\newblock Cancer. 1961;14:1272--1294.

\bibitem{Spratt:1964}
Spratt JS, Spratt TL.
\newblock Rates of Growth of Pulmonary Metastases and Host Survival.
\newblock Annals of Surgery. 1964;159(2):161--171.

\bibitem{Jones:2008}
Jones S, Chen Wd, Parmigiani G, Diehl F, Beerenwinkel N, Antal T, et~al.
\newblock Comparative lesion sequencing provides insights into tumor evolution.
\newblock Proceedings of the National Academy of Sciences.
  2008;105(11):4283--4288.
\newblock doi:{10.1073/pnas.0712345105}.

\bibitem{Haustermans:1998}
Haustermans K, Fowler J, Geboes K, Christiaens MR, Lerut A, van~der Schueren E.
\newblock Relationship between potential doubling time (Tpot), labeling index
  and duration of {DNA} synthesis in 60 esophageal and 35 breast tumors: is it
  worthwhile to measure Tpot?
\newblock Radiotherapy and Oncology. 1998;46(2):157--167.
\newblock doi:{10.1016/s0167-8140(97)00164-3}.

\bibitem{Denekamp:1992}
Denekamp J.
\newblock New Approaches to the Measurement of Proliferation Rates.
\newblock In: Maragoudakis ME, Gullino P, Lelkes PI, editors. Angiogenesis in
  Health and Disease. Boston, MA: Springer US; 1992. p. 333--337.

\bibitem{Bertuzzi:1997}
Bertuzzi A, Gandolfi A, Sinisgalli C, Starace G, Ubezio P.
\newblock Cell loss and the concept of potential doubling time.
\newblock Cytometry. 1997;29(1):34--40.
\newblock doi:{10.1002/(sici)1097-0320(19970901)29:1<34::aid-cyto3>3.0.co;2-d}.

\bibitem{Fournier:1980}
von Fournier D, Weber E, Hoeffken W, Bauer M, Kubli F, Barth V.
\newblock Growth rate of 147 mammary carcinomas.
\newblock Cancer. 1980;45:2198--2207.

\bibitem{Kuroishi:1990}
Kuroishi T, Tominaga S, Morimoto T, Tashiro H, Itoh S, Watanabe H, et~al.
\newblock Tumor Growth Rate and Prognosis of Breast Cancer Mainly Detected by
  Mass Screening.
\newblock Japanese Journal of Cancer Research. 1990;81(5):454--462.
\newblock doi:{10.1111/j.1349-7006.1990.tb02591.x}.

\bibitem{Peer:1993}
Peer PGM, Van~Dijck JAAM, Verbeek ALM, Hendriks JHCL, Holland R.
\newblock Age-dependent growth rate of primary breast cancer.
\newblock Cancer. 1993;71(11):3547--3551.
\newblock
  doi:{10.1002/1097-0142(19930601)71:11<3547::aid-cncr2820711114>3.0.co;2-c}.

\bibitem{Ryu:2014}
Ryu EB, Chang JM, Seo M, Kim SA, Lim JH, Moon WK.
\newblock Tumour volume doubling time of molecular breast cancer subtypes
  assessed by serial breast ultrasound.
\newblock European Radiology. 2014;24(9):2227--2235.
\newblock doi:{10.1007/s00330-014-3256-0}.

\bibitem{Foernvik:2015}
F\"ornvik D, L{\aa}ng K, Andersson I, Dustler M, Borgquist S, Timberg P.
\newblock Estimates of breast cancer growth rate from mammograms and its
  relation to tumour characteristics.
\newblock Radiation Protection Dosimetry. 2015;169(1-4):151--157.
\newblock doi:{10.1093/rpd/ncv417}.

\bibitem{Zhang:2017}
Zhang S, Ding Y, Zhou Q, Wang C, Wu P, Dong J.
\newblock Correlation Factors Analysis of Breast Cancer Tumor Volume Doubling
  Time Measured by 3D-Ultrasound.
\newblock Medical Science Monitor. 2017;23:3147--3153.
\newblock doi:{10.12659/msm.901566}.

\bibitem{Kusama:1972}
Kusama S, Spratt~Jr JS, Donegan WL, Watson FR, Cunningham C.
\newblock The gross rates of growth of human mammary carcinoma.
\newblock Cancer. 1972;30(2):594--599.
\newblock
  doi:{10.1002/1097-0142(197208)30:2<594::aid-cncr2820300241>3.0.co;2-2}.

\bibitem{Friberg:1997}
Friberg S, Mattson S.
\newblock On the growth rates of human malignant tumors: implications for
  medical decision making.
\newblock Journal of surgical oncology. 1997;65:284--297.

\bibitem{Awwad:2013}
Awwad H.
\newblock Radiation Oncology: Radiobiological and Physiological Perspectives.
\newblock Springer Netherlands; 2013.

\bibitem{Zabicki:2006}
Zabicki K, Colbert JA, Dominguez FJ, Gadd MA, Hughes KS, Jones JL, et~al.
\newblock Breast Cancer Diagnosis in Women $\leq$ 40 versus 50 to 60 Years:
  Increasing Size and Stage Disparity Compared With Older Women Over Time.
\newblock Annals of Surgical Oncology. 2006;13(8):1072--1077.
\newblock doi:{10.1245/aso.2006.03.055}.

\bibitem{Lee:2016}
Lee SH, Kim YS, Han W, Ryu HS, Chang JM, Cho N, et~al.
\newblock Tumor growth rate of invasive breast cancers during wait times for
  surgery assessed by ultrasonography.
\newblock Medicine. 2016;95(37):e4874.
\newblock doi:{10.1097/md.0000000000004874}.

\bibitem{DeLAulnoit:2018}
de~l'Aulnoit AH, Rogoz B, Pin{\c{c}}on C, de~l'Aulnoit DH.
\newblock Metastasis-free interval in breast cancer patients: thirty-year
  trends and time dependency of prognostic factors. A retrospective analysis
  based on a single institution experience.
\newblock The Breast. 2018;37:80--88.
\newblock doi:{10.1016/j.breast.2017.10.008}.

\bibitem{Bolin:1983}
Bolin S, Nilsson E, Sj\"odahl R. Carcinoma of the colon and rectum--growth
  rate.; 1983.

\bibitem{Tada:1984}
Tada M, Misaki F, Kawai K.
\newblock Growth rates of colorectal carcinoma and adenoma by roentgenologic
  follow-up observations.
\newblock Gastroenterologia Japonica. 1984;19:550--555.

\bibitem{Choi:2013}
Choi SJ, Kim HS, Ahn SJ, Jeong YM, Choi HY.
\newblock Evaluation of the growth pattern of carcinoma of colon and rectum by
  {MDCT}.
\newblock Acta Radiologica. 2013;54(5):487--492.
\newblock doi:{10.1177/0284185113475923}.

\bibitem{Finlay:1988}
Finlay IG, Meek D, Bruntont F, McArdle CS.
\newblock Growth rate of hepatic metastases in colorectal carcinoma.
\newblock British Journal of Surgery. 1988;75(7):641--644.
\newblock doi:{10.1002/bjs.1800750707}.

\bibitem{Tanaka:2004}
Tanaka K, Shimada H, Miura M, Fujii Y, Yamaguchi S, Endo I, et~al.
\newblock Metastatic Tumor Doubling Time: Most Important Prehepatectomy
  Predictor of Survival and Nonrecurrence of Hepatic Colorectal Cancer
  Metastasis.
\newblock World Journal of Surgery. 2004;28(3):263--270.
\newblock doi:{10.1007/s00268-003-7088-3}.

\bibitem{Tomimaru:2008}
Tomimaru Y, Noura S, Ohue M, Okami J, Oda K, Higashiyama M, et~al.
\newblock Metastatic Tumor Doubling Time Is an Independent Predictor of
  Intrapulmonary Recurrence after Pulmonary Resection of Solitary Pulmonary
  Metastasis from Colorectal Cancer.
\newblock Digestive Surgery. 2008;25(3):220--225.
\newblock doi:{10.1159/000140693}.

\bibitem{Wilson:1993}
Wilson MS, West CM, Wilson GD, Roberts SA, James RD, Schofield PF.
\newblock Intra-tumoral heterogeneity of tumour potential doubling times (Tpot)
  in colorectal cancer.
\newblock British journal of cancer. 1993;68:501--506.

\bibitem{Kornprat:2011}
Kornprat P, Pollheimer MJ, Lindtner RA, Schlemmer A, Rehak P, Langner C.
\newblock Value of Tumor Size as a Prognostic Variable in Colorectal Cancer.
\newblock American Journal of Clinical Oncology. 2011;34(1):43--49.
\newblock doi:{10.1097/coc.0b013e3181cae8dd}.

\bibitem{Ding:2017}
Ding Z, Wang Z, Huang S, Zhong S, Lin J.
\newblock Comparison of laparoscopic vs. open surgery for rectal cancer.
\newblock Molecular and Clinical Oncology. 2017;6(2):170--176.
\newblock doi:{10.3892/mco.2016.1112}.

\bibitem{Waaijer:2003}
Waaijer A, Terhaard CHJ, Dehnad H, Hordijk GJ, van Leeuwen MS, Raaymakers CPJ,
  et~al.
\newblock Waiting times for radiotherapy: consequences of volume increase for
  the {TCP} in oropharyngeal carcinoma.
\newblock Radiotherapy and Oncology. 2003;66(3):271--276.
\newblock doi:{10.1016/s0167-8140(03)00036-7}.

\bibitem{Jensen:2007}
Jensen AR, Nellemann HM, Overgaard J.
\newblock Tumor progression in waiting time for radiotherapy in head and neck
  cancer.
\newblock Radiotherapy and Oncology. 2007;84(1):5--10.
\newblock doi:{10.1016/j.radonc.2007.04.001}.

\bibitem{Galante:1982}
Galante E, Gallus G, Chiesa F, Bono A, Bettoni I, Molinari R.
\newblock Growth rate of head and neck tumors.
\newblock European Journal of Cancer and Clinical Oncology.
  1982;18(8):707--712.
\newblock doi:{10.1016/0277-5379(82)90067-0}.

\bibitem{Umino:1997}
Umino S, Hayashi S, Ono S.
\newblock Doubling time of pulmonary metastases of adenoid cystic carcinoma.
\newblock International Journal of Oral and Maxillofacial Surgery. 1997;26:48.
\newblock doi:{10.1016/s0901-5027(97)80987-3}.

\bibitem{Zackrisson:1997}
Zackrisson B, Gustafsson H, Stenling R, Flygare P, Wilson GD.
\newblock Predictive value of potential doubling time in head and neck cancer
  patients treated by conventional radiotherapy.
\newblock International Journal of Radiation
  Oncology$\ast$Biology$\ast$Physics. 1997;38(4):677--683.
\newblock doi:{10.1016/s0360-3016(97)00066-7}.

\bibitem{Muto:2004}
Muto M, Nakane M, Katada C, Sano Y, Ohtsu A, Esumi H, et~al.
\newblock Squamous cell carcinoma in situ at oropharyngeal and hypopharyngeal
  mucosal sites.
\newblock Cancer. 2004;101(6):1375--1381.
\newblock doi:{10.1002/cncr.20482}.

\bibitem{Markou:2011}
Markou K, Goudakos J, Triaridis S, Konstantinidis J, Vital V, Nikolaou A.
\newblock The role of tumor size and patient's age as prognostic factors in
  laryngeal cancer.
\newblock Hippokratia. 2011;15(21607041):75--80.

\bibitem{Kerr:1984}
Kerr KM, Lamb D.
\newblock Actual growth rate and tumour cell proliferation in human pulmonary
  neoplasms.
\newblock British Journal Of Cancer. 1984;50:343.

\bibitem{Arai:1994}
Arai T, Kuroishi T, Saito Y, Kurita Y, Naruke T, Kaneko M.
\newblock Tumor Doubling Time and Prognosis in Lung Cancer Patients: Evaluation
  from Chest Films and Clinical Follow-up Study.
\newblock Japanese Journal of Clinical Oncology.
  1994;doi:{10.1093/oxfordjournals.jjco.a039706}.

\bibitem{Detterbeck:2008}
Detterbeck FC, Gibson CJ.
\newblock Turning Gray: The Natural History of Lung Cancer Over Time.
\newblock Journal of Thoracic Oncology. 2008;3(7):781--792.
\newblock doi:{10.1097/jto.0b013e31817c9230}.

\bibitem{Henschke:2012}
Henschke CI, Yankelevitz DF, Yip R, Reeves AP, Farooqi A, Xu D, et~al.
\newblock Lung Cancers Diagnosed at Annual {CT} Screening: Volume Doubling
  Times.
\newblock Radiology. 2012;263(2):578--583.
\newblock doi:{10.1148/radiol.12102489}.

\bibitem{Yoo:2008}
Yoo H, Nam BH, Yang HS, Shin SH, Lee JS, Lee SH.
\newblock Growth rates of metastatic brain tumors in nonsmall cell lung cancer.
\newblock Cancer. 2008;113(5):1043--1047.
\newblock doi:{10.1002/cncr.23676}.

\bibitem{Fowler:2001}
Fowler JF.
\newblock Biological Factors Influencing Optimum Fractionation in Radiation
  Therapy.
\newblock Acta Oncologica. 2001;40(6):712--717.
\newblock doi:{10.1080/02841860152619124}.

\bibitem{Bando:2002}
Bando T.
\newblock A new method of segmental resection for primary lung cancer:
  intermediate results.
\newblock European Journal of Cardio-Thoracic Surgery. 2002;21(5):894--899.
\newblock doi:{10.1016/s1010-7940(02)00122-7}.

\bibitem{Strand:2006}
Strand TE.
\newblock Survival after resection for primary lung cancer: a population based
  study of 3211 resected patients.
\newblock Thorax. 2006;61(8):710--715.
\newblock doi:{10.1136/thx.2005.056481}.

\bibitem{DAmico:1993}
D'Amico AV, Hanks GE.
\newblock Linear regressive analysis using prostate-specific antigen doubling
  time for predicting tumor biology and clinical outcome in prostate cancer.
\newblock Cancer. 1993;72:2638--2643.

\bibitem{Werahera:2011}
Werahera PN, Glode LM, Rosa FGL, Lucia MS, Crawford ED, Easterday K, et~al.
\newblock Proliferative Tumor Doubling Times of Prostatic Carcinoma.
\newblock Prostate Cancer. 2011;2011:1--7.
\newblock doi:{10.1155/2011/301850}.

\bibitem{Zharinov:2017}
Zharinov GM, Bogomolov OA, Neklasova NN, Anisimov VN.
\newblock Pretreatment prostate specific antigen doubling time as prognostic
  factor in prostate cancer patients.
\newblock Oncoscience. 2017;4:7--13.
\newblock doi:{10.18632/oncoscience.337}.

\bibitem{Berges:1995}
Berges RR, Vukanovic J, Epstein JI, CarMichel M, Cisek L, Johnson DE, et~al.
\newblock Implication of cell kinetic changes during the progression of human
  prostatic cancer.
\newblock Clinical cancer research : an official journal of the American
  Association for Cancer Research. 1995;1:473--480.

\bibitem{Haustermans:1997}
Haustermans KMG, Hofland I, Poppel HV, Oyen R, de~Voorde WV, Begg AC, et~al.
\newblock Cell kinetic measurements in prostate cancer.
\newblock International Journal of Radiation
  Oncology$\ast$Biology$\ast$Physics. 1997;37(5):1067--1070.
\newblock doi:{10.1016/s0360-3016(96)00579-2}.

\bibitem{Renshaw:1999}
Renshaw AA, Richie JP, Loughlin KR, Jiroutek M, Chung A, D'Amico AV.
\newblock Maximum diameter of prostatic carcinoma is a simple, inexpensive, and
  independent predictor of prostate-specific antigen failure in radical
  prostatectomy specimens. Validation in a cohort of 434 patients.
\newblock American journal of clinical pathology. 1999;111:641--644.

\bibitem{Johnson:2013}
Johnson SB, Hamstra DA, Jackson WC, Zhou J, Foster B, Foster C, et~al.
\newblock Larger Maximum Tumor Diameter at Radical Prostatectomy Is Associated
  With Increased Biochemical Failure, Metastasis, and Death From Prostate
  Cancer After Salvage Radiation for Prostate Cancer.
\newblock International Journal of Radiation
  Oncology$\ast$Biology$\ast$Physics. 2013;87(2):275--281.
\newblock doi:{10.1016/j.ijrobp.2013.05.043}.

\bibitem{Serres:2012}
Serres S, Soto MS, Hamilton A, McAteer MA, Carbonell WS, Robson MD, et~al.
\newblock Molecular {MRI} enables early and sensitive detection of brain
  metastases.
\newblock Proceedings of the National Academy of Sciences.
  2012;109(17):6674--6679.
\newblock doi:{10.1073/pnas.1117412109}.

\bibitem{Fujiwara:2015}
Fujiwara S, Yao K, Nagahama T, Uchita K, Kanemitsu T, Tsurumi K, et~al.
\newblock Can we accurately diagnose minute gastric cancers ($\le5$mm)?
  Chromoendoscopy (CE) vs magnifying endoscopy with narrow band imaging
  (M-NBI).
\newblock Gastric Cancer. 2015;18(3):590--596.

\bibitem{Wang:2017}
Wang L.
\newblock Early Diagnosis of Breast Cancer.
\newblock Sensors. 2017;17(7):1572.
\newblock doi:{10.3390/s17071572}.

\bibitem{Chignola:2005}
Chignola R, Foroni RI.
\newblock Estimating the Growth Kinetics of Experimental Tumors From as Few as
  Two Determinations of Tumor Size: Implications for Clinical Oncology.
\newblock {IEEE} Transactions on Biomedical Engineering. 2005;52(5):808--815.
\newblock doi:{10.1109/tbme.2005.845219}.

\bibitem{Fillon:2018}
Fillon M.
\newblock Better Guidelines Needed for Cancer Survivorship Management.
\newblock {CA}: A Cancer Journal for Clinicians. 2018;68(6):392--393.
\newblock doi:{10.3322/caac.21435}.

\bibitem{Lee:2006}
Lee SP, Sun JR, Qian H, McBride WH, Withers HR.
\newblock Characterization of Metastatic Tumor Formation by the Colony Size
  Distribution.
\newblock arXiv pre-print. 2006;.

\bibitem{Schmid:1993}
Schmid HP, McNeal JE, Stamey TA.
\newblock Observations on the doubling time of prostate cancer. The use of
  serial prostate-specific antigen in patients with untreated disease as a
  measure of increasing cancer volume.
\newblock Cancer. 1993;71(6):2031--2040.
\newblock
  doi:{10.1002/1097-0142(19930315)71:6<2031::aid-cncr2820710618>3.0.co;2-q}.

\bibitem{Andre:2004}
Andre F, Slimane K, Bachelot T, Dunant A, Namer M, Barrelier A, et~al.
\newblock Breast Cancer With Synchronous Metastases: Trends in Survival During
  a 14-Year Period.
\newblock Journal of Clinical Oncology. 2004;22(16):3302--3308.
\newblock doi:{10.1200/jco.2004.08.095}.

\bibitem{Boutros:2015}
Boutros C, Mazouni C, Lerebours F, Stevens D, Lei X, Gonzalez-Angulo AM, et~al.
\newblock A preoperative nomogram to predict the risk of synchronous distant
  metastases at diagnosis of primary breast cancer.
\newblock British Journal of Cancer. 2015;112(6):992--997.
\newblock doi:{10.1038/bjc.2015.34}.

\bibitem{Yilmaz:2018}
Yilmaz U, Marks LB.
\newblock Estimating changes in the rate of synchronous and metachronous
  metastases over time: Analysis of {SEER} data.
\newblock Advances in Radiation Oncology. 2018;3(1):70--75.
\newblock doi:{10.1016/j.adro.2017.09.007}.

\bibitem{Kim:2013}
Kim H, Choi DH, Park W, Huh SJ, Nam SJ, Lee JE, et~al.
\newblock Prognostic factors for survivals from first relapse in breast cancer
  patients: analysis of deceased patients.
\newblock Radiation Oncology Journal. 2013;31(4):222.
\newblock doi:{10.3857/roj.2013.31.4.222}.

\bibitem{Fitzpatrick:2013}
Fitzpatrick DJ, Lai CS, Parkyn RF, Walters D, Humeniuk V, Walsh DCA.
\newblock Time to Breast Cancer Relapse Predicted By Primary Tumour
  Characteristics, Not Lymph Node Involvement.
\newblock World Journal of Surgery. 2013;38(7):1668--1675.
\newblock doi:{10.1007/s00268-013-2397-7}.

\bibitem{Nowikiewicz:2015}
Nowikiewicz T, Wi{\'{s}}niewska M, Wi{\'{s}}niewski M, Biedka M, G{\l}owacka I,
  Kozak D, et~al.
\newblock Overall survival and disease-free survival in breast cancer patients
  treated at the Oncology Centre in Bydgoszcz {\textendash} analysis of more
  than six years of follow-up.
\newblock Wsp{\'{o}}{\l}czesna Onkologia. 2015;4:284--289.
\newblock doi:{10.5114/wo.2015.54387}.

\bibitem{Kemeny:2002}
Kemeny MM, Adak S, Gray B, Macdonald JS, Smith T, Lipsitz S, et~al.
\newblock Combined-Modality Treatment for Resectable Metastatic Colorectal
  Carcinoma to the Liver: Surgical Resection of Hepatic Metastases in
  Combination With Continuous Infusion of Chemotherapy - An Intergroup Study.
\newblock Journal of Clinical Oncology. 2002;20(6):1499--1505.
\newblock doi:{10.1200/JCO.2002.20.6.1499}.

\bibitem{Park:2013}
Park JH, Kim TY, Lee KH, Han SW, Oh DY, Im SA, et~al.
\newblock The beneficial effect of palliative resection in metastatic
  colorectal cancer.
\newblock British Journal Of Cancer. 2013;108:1425.

\bibitem{Lykoudis:2014}
Lykoudis PM, O'Reilly D, Nastos K, Fusai G.
\newblock Systematic review of surgical management of synchronous colorectal
  liver metastases.
\newblock Br J Surg. 2014;101(6):605--612.

\bibitem{Elferink:2015}
Elferink MAG, de~Jong KP, Klaase JM, Siemerink EJ, de~Wilt JHW.
\newblock Metachronous metastases from colorectal cancer: a population-based
  study in North-East Netherlands.
\newblock International Journal of Colorectal Disease. 2015;30(2):205--212.
\newblock doi:{10.1007/s00384-014-2085-6}.

\bibitem{Holch:2017}
Holch JW, Demmer M, Lamersdorf C, Michl M, Schulz C, von Einem JC, et~al.
\newblock Pattern and Dynamics of Distant Metastases in Metastatic Colorectal
  Cancer.
\newblock Visceral Medicine. 2017;33(1):70--75.
\newblock doi:{10.1159/000454687}.

\bibitem{Hohenberger:1994}
Hohenberger P, Schlag PM, Gerneth T, Herfarth C.
\newblock Pre- and postoperative carcinoembryonic antigen determinations in
  hepatic resection for colorectal metastases. Predictive value and
  implications for adjuvant treatment based on multivariate analysis.
\newblock Annals of Surgery. 1994;219:135--143.

\bibitem{Nordlinger:2009}
Nordlinger B, Van~Cutsem E, Gruenberger T, Glimelius B, Poston G, Rougier P,
  et~al.
\newblock Combination of surgery and chemotherapy and the role of targeted
  agents in the treatment of patients with colorectal liver metastases:
  recommendations from an expert panel.
\newblock Annals of Oncology. 2009;20(6):985--992.
\newblock doi:{10.1093/annonc/mdn735}.

\bibitem{Sturesson:2017}
Sturesson C, Valdimarsson VT, Blomstrand E, Eriksson S, Nilsson JH, Syk I,
  et~al.
\newblock Liver-first strategy for synchronous colorectal liver metastases - an intention-to-treat analysis.
\newblock HPB. 2017;19(1):52--58.
\newblock doi:{https://doi.org/10.1016/j.hpb.2016.10.005}.

\bibitem{Ferlito:2001}
Ferlito A, Shaha AR, Silver CE, Rinaldo A, Mondin V.
\newblock Incidence and Sites of Distant Metastases from Head and Neck Cancer.
\newblock {ORL}. 2001;63(4):202--207.
\newblock doi:{10.1159/000055740}.

\bibitem{Jain:2013}
Jain KS, Sikora AG, Baxi SS, Morris LGT.
\newblock Synchronous cancers in patients with head and neck cancer.
\newblock Cancer. 2013;119(10):1832--1837.
\newblock doi:{10.1002/cncr.27988}.

\bibitem{Liu:2007}
Liu SA, Wong YK, Lin JC, Poon CK, Tung KC, Tsai WC.
\newblock Impact of recurrence interval on survival of oral cavity squamous
  cell carcinoma patients after local relapse.
\newblock Otolaryngology-Head and Neck Surgery. 2007;136(1):112--118.
\newblock doi:{10.1016/j.otohns.2006.07.002}.

\bibitem{Ebrahimi:2012}
Ebrahimi A, Clark JR, Ahmadi N, Palme CE, Morgan GJ, Veness MJ.
\newblock Prognostic significance of disease-free interval in head and neck
  cutaneous squamous cell carcinoma with nodal metastases.
\newblock Head {\&} Neck. 2012;35(8):1138--1143.
\newblock doi:{10.1002/hed.23096}.

\bibitem{Wiegand:2015}
Wiegand S, Zimmermann A, Wilhelm T, Werner JA.
\newblock Survival After Distant Metastasis in Head and Neck Cancer.
\newblock Anticancer research. 2015;35:5499--5502.

\bibitem{Toennies:2014}
T\"onnies M, Pfannschmidt J, Bauer TT, Kollmeier J, T\"onnies S, Kaiser D.
\newblock Metastasectomy for Synchronous Solitary Non-Small Cell Lung Cancer
  Metastases.
\newblock The Annals of Thoracic Surgery. 2014;98(1):249--256.
\newblock doi:{10.1016/j.athoracsur.2014.03.028}.

\bibitem{al-Kattan:1997}
al~Kattan K, Sepsas E, Fountain SW, Townsend ER.
\newblock Disease recurrence after resection for stage I lung cancer.
\newblock European journal of cardio-thoracic surgery : official journal of the
  European Association for Cardio-thoracic Surgery. 1997;12:380--384.

\bibitem{Hung:2010}
Hung JJ, Jeng WJ, Hsu WH, Wu KJ, Chou TY, Hsieh CC, et~al.
\newblock Prognostic factors of postrecurrence survival in completely resected
  stage I non-small cell lung cancer with distant metastasis.
\newblock Thorax. 2010;65(3):241--245.
\newblock doi:{10.1136/thx.2008.110825}.

\bibitem{Farsi:2017}
Farsi AA, Swaminath A, Ellis P.
\newblock Patterns of Relapse in Small Cell Lung Cancer ({SCLC}): A
  Retrospective Analysis of Outcomes from a Single Canadian Center.
\newblock Journal of Thoracic Oncology. 2017;12(1):S727--S728.
\newblock doi:{10.1016/j.jtho.2016.11.962}.

\bibitem{Koo:2015}
Koo KC, Yoo H, Kim KH, Park SU, Han KS, Rha KH, et~al.
\newblock Prognostic Impact of Synchronous Second Primary Malignancies on the
  Overall Survival of Patients with Metastatic Prostate Cancer.
\newblock Journal of Urology. 2015;193(4):1239--1244.
\newblock doi:{10.1016/j.juro.2014.10.088}.

\bibitem{Fontenot:2017}
Fontenot PA Jr, Nehra A, Parker W, Wyre H, Mirza M, Duchene DA, et~al.
\newblock Metastatic prostate cancer in the modern era of {PSA} screening.
\newblock International braz j urol. 2017;43(3):416--421.
\newblock doi:{10.1590/s1677-5538.ibju.2016.0340}.

\bibitem{Almeida:2018}
Almeida PL, Pereira BJ.
\newblock Local treatment of metastatic prostate cancer: what is the evidence
  so far?
\newblock Prostate Cancer. 2018;2018:1--7.
\newblock doi:{10.1155/2018/2654572}.

\bibitem{Boorjian:2011}
Boorjian SA, Thompson RH, Tollefson MK, Rangel LJ, Bergstralh EJ, Blute ML,
  et~al.
\newblock Long-Term Risk of Clinical Progression After Biochemical Recurrence
  Following Radical Prostatectomy: The Impact of Time from Surgery to
  Recurrence.
\newblock European Urology. 2011;59(6):893--899.
\newblock doi:{10.1016/j.eururo.2011.02.026}.

\bibitem{Toussi:2016}
Toussi A, Stewart-Merrill SB, Boorjian SA, Psutka SP, Thompson RH, Frank I,
  et~al.
\newblock Standardizing the Definition of Biochemical Recurrence after Radical
  Prostatectomy{\textemdash}What Prostate Specific Antigen Cut Point Best
  Predicts a Durable Increase and Subsequent Systemic Progression?
\newblock Journal of Urology. 2016;195(6):1754--1759.
\newblock doi:{10.1016/j.juro.2015.12.075}.

\bibitem{Obenauf:2015}
Obenauf AC, Massagu{\'{e}} J.
\newblock Surviving at a Distance: Organ-Specific Metastasis.
\newblock Trends in Cancer. 2015;1(1):76--91.
\newblock doi:{10.1016/j.trecan.2015.07.009}.

\bibitem{Durrett:2015}
Durrett R.
\newblock Branching process models of cancer. vol. 1.1 of Stochastics in
  biological systems.
\newblock 1st ed. Springer International Publishing; 2015.

\bibitem{Haeno:2012}
Haeno H, Gonen M, Davis MB, Herman JM, Iacobuzio-Donahue CA, Michor F.
\newblock Computational Modeling of Pancreatic Cancer Reveals Kinetics of
  Metastasis Suggesting Optimum Treatment Strategies.
\newblock Cell. 2012;148(1-2):362--375.
\newblock doi:{10.1016/j.cell.2011.11.060}.

\bibitem{Yachida:2010}
Yachida S, Jones S, Bozic I, Antal T, Leary R, Fu B, et~al.
\newblock Distant metastasis occurs late during the genetic evolution of
  pancreatic cancer.
\newblock Nature. 2010;467:1114.

\bibitem{Nicholson:2019}
Nicholson MD, Antal T.
\newblock Competing evolutionary paths in growing populations with applications
  to multidrug resistance.
\newblock {PLOS} Computational Biology. 2019;15(4):e1006866.
\newblock doi:{10.1371/journal.pcbi.1006866}.

\bibitem{Gundem:2015}
Gundem G, , Loo PV, Kremeyer B, Alexandrov LB, Tubio JMC, et~al.
\newblock The evolutionary history of lethal metastatic prostate cancer.
\newblock Nature. 2015;520(7547):353--357.
\newblock doi:{10.1038/nature14347}.

\bibitem{Bozic:2010}
Bozic I, Antal T, Ohtsuki H, Carter H, Kim D, Chen S, et~al.
\newblock Accumulation of driver and passenger mutations during tumor
  progression.
\newblock Proceedings of the National Academy of Sciences.
  2010;107(43):18545--18550.
\newblock doi:{10.1073/pnas.1010978107}.

\bibitem{Williams:1965a}
Williams T.
\newblock The Basic Birth-Death Model for Microbial Infections.
\newblock Journal of the Royal Statistical Society Series B (Methodological).
  1965;27(2):338--360.

\bibitem{Waugh:1972c}
Waugh WAO.
\newblock Uses of the sojourn time series for Markovian birth process.
\newblock In: Proceedings of the Sixth Berkeley Symposium on Mathematical
  Statistics and Probability, Volume 3: Probability Theory. Berkeley, CA:
  University of California Press; 1972. p. 501--514.

\end{thebibliography}

%
%
%
%

\end{document}